\documentclass[showpacs,superscriptaddress,floatfix,amsmath,amssymb,twocolumn,pra]{revtex4-1}
\usepackage{bm}
\usepackage{graphicx}
\usepackage{color}
\usepackage[normalem]{ulem}
\usepackage{enumerate}

\def\sgn{{\rm sgn}}

\def\Re{{\rm Re}}
\def\Im{{\rm Im}}

\newcommand{\Av}[1]{\langle #1 \rangle}
\newcommand{\DblAv}[1]{\langle\langle #1 \rangle\rangle}

\begin{document}

\title {Interplay of driving and frequency noise in the spectra of vibrational systems
}
\author{Yaxing Zhang}
\affiliation{Department of Physics and Astronomy, Michigan State University, East Lansing, MI 48824, USA}
\author{J. Moser}
\affiliation{ICFO - Institut de Ciencies Fotoniques,
Mediterranean Technology Park, 08860 Castelldefels, Barcelona,
Spain}
\author{J. G\"uttinger}
\affiliation{ICFO - Institut de Ciencies Fotoniques,
Mediterranean Technology Park, 08860 Castelldefels, Barcelona,
Spain}
\author{A. Bachtold}
\affiliation{ICFO - Institut de Ciencies Fotoniques,
Mediterranean Technology Park, 08860 Castelldefels, Barcelona,
Spain}
\author{M. I. Dykman}
\affiliation{Department of Physics and Astronomy, Michigan State University, East Lansing, MI 48824, USA}

\date{\today}

\begin{abstract}
We study the spectral effect of the fluctuations of the vibration frequency. Such fluctuations play a major role in nanomechanical and other mesoscopic vibrational systems. We find that, for periodically driven systems, the interplay of the driving and frequency fluctuations results in specific spectral features. We present measurements on a carbon nanotube resonator and show that our theory allows not only the characterization of the frequency fluctuations but also the quantification of the decay rate without ring-down measurements. The results bear on identifying the decoherence of mesoscopic oscillators and on the general problem of resonance fluorescence and light scattering by oscillators. 

\end{abstract}


\maketitle

The spectrum of response and the power spectrum of an oscillator is a textbook problem that goes back to Lorentz and Einstein \cite{Lorentz1916,Einstein1910b,Heitler2010}.   It has attracted much attention recently in the context of nanomechanical systems. Here, the spectra are a major source of information about the classical and quantum dynamics \cite{Cleland2002,Sazonova2004,Lifshitz2008,Steele2009,Lassagne2009,Fong2012,Siria2012,Gavartin2013}. This is the case also for mesoscopic oscillators of different nature, such as superconducting cavity modes \cite{Gao2007,Neill2013,Burnett2013a,Faoro2014} and optomechanical systems \cite{Aspelmeyer2013}.
Mesoscopic oscillators  experience comparatively large fluctuations. Along with dissipation, these fluctuations determine the shape of the vibrational spectra.  

A well-understood and most frequently considered \cite{Heitler2010} source of fluctuations is thermal noise that comes from the coupling of an oscillator (vibrational system) to a thermal reservoir and is related to dissipation by the fluctuation-dissipation theorem. Dissipation leads to the broadening of the oscillator power spectrum and the spectrum of the response to external driving. 

Spectral broadening can also come from fluctuations of the oscillator frequency, which play an important role in mesoscopic oscillators. For nanomechanical resonators, frequency fluctuations can be caused by tension and mass fluctuations, fluctuations of the charge in the substrate, or dispersive intermode coupling \cite{Steele2009,Lassagne2009,Fong2012,Siria2012,Gavartin2013,Dykman2010,Yang2011,Barnard2012,Miao2014}, whereas for electromagnetic cavity modes they can come from fluctuations of the effective dielectric constant \cite{Gao2007,Neill2013}. Identifying different broadening mechanisms  is a delicate task that has been attracting much attention \cite{Fong2012,Gavartin2013,Gao2007,Barnard2012,Eichler2013,Meenehan2014}.

In this paper we study the combined effect of periodic driving and frequency fluctuations on the power spectra of nanomechanical  vibrational systems. For a linear oscillator with no frequency fluctuations, driving leads to a $\delta$-like peak at the driving frequency $\omega_F$ \cite{Heitler2010,Cleland2002}, because here the only effect of the driving is forced vibrations linearly superimposed on thermal motion. Frequency fluctuations make forced vibrations random. As we show, this qualitatively changes the spectrum leading to characteristic new spectral features.  We observe these features in a carbon-nanotube resonator and use them to separate the energy relaxation rate from the overall broadening of the power spectrum in the absence of driving, as well as reveal and explore the narrow-band frequency noise. We predict that for fluctuating nonlinear vibrational systems, too,  even weak driving leads to a very specific extra spectral structure.
\begin{figure}[h]
\includegraphics
{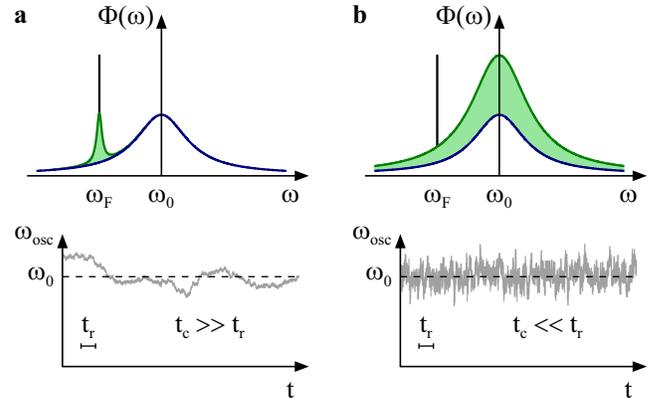}
\caption{Top: sketches of the power spectra of a driven linear oscillator $\Phi(\omega)$. Panels (a) and (b) refer  to large  and small correlation time of the frequency noise $t_c$ compared to the oscillator relaxation time $t_r$, respectively, i.e., to narrow- and broad-band frequency noise.   The blue (lower) line shows the spectrum of thermal fluctuations in the absence of driving; it is centered at the oscillator eigenfrequency $\omega_0=\langle \omega_{\rm osc}(t)\rangle$. In the presence of driving there is added a $\delta$-peak at the driving frequency $\omega_F$.  The green areas show the spectral features from the interplay of the driving and fluctuations of $\omega_{\rm osc}(t)$. Bottom panels: $\omega_{\rm osc}(t)$ for $t_c\gg t_r$  (a) and $t_c\ll t_r$ (b).}
\label{fig:sketch}
\end{figure}

For a linear oscillator, the spectral features resulting from the interplay of driving and frequency noise are sketched in Fig.~\ref{fig:sketch}. The two limiting cases shown in Fig.~\ref{fig:sketch} correspond to the long and short correlation time of the frequency noise $t_c$ compared to the oscillator relaxation (decay) time $t_r$. For $t_c\gg t_r$ (panel a) the oscillator frequency $\omega_{\rm osc}(t)$ slowly fluctuates about what can be called the eigenfrequency $\omega_0=\langle \omega_{\rm osc}(t)\rangle$. One can then think of slow fluctuations of the oscillator susceptibility $\chi$, which depends on the detuning of the driving frequency $\omega_F$ from $\omega_{\rm osc}(t)$. The associated slow fluctuations of the amplitude and phase of forced vibrations at frequency $\omega_F$ lead to a finite-width spectral peak centered at $\omega_F$. This is a frequency-domain analog of the Einstein light scattering due to spatial susceptibility fluctuations \cite{Einstein1910}.

For $t_c \ll t_r$ (panel b), driving-induced random vibrations quickly lose the memory of the driving frequency. They become similar to thermal vibrations. However, their amplitude is determined by the driving, not the temperature. This leads to a spectral peak centered at the oscillator eigenfrequency  $\omega_0$, with the height quadratic in the driving amplitude. 

In the quantum picture, one can think that, as a result of pumping by a driving field, the oscillator emits energy quanta. For the familiar example of an oscillating charge driven by an electromagnetic field these quanta can be photons, and one can speak of light scattering and fluorescence by an oscillator. A quantum is emitted over time $t_r$ after the absorption event. For $t_c\ll t_r$ the frequency of the quantum is uncorrelated with the excitation frequency $\omega_F$. 
This is a fluorescence-type process. The energy difference $\hbar(\omega_F-\omega_0)$ comes from the frequency noise. For $t_c\gg t_r$ emission occurs at frequencies close to $\omega_F$. In the both cases the spectrum is qualitatively different from just a $\delta$-like peak in the absence of frequency fluctuations \cite{Heitler2010}. 

To describe the power spectrum of a nonlinear system one has to go beyond the approximation implied above, where only the linear susceptibility is fluctuating. If driving is described by the term $-qF(t)$ in the oscillator Hamiltonian, where $q$ is the oscillator coordinate and $F(t)=F\cos\omega_Ft$ is the driving force, to obtain terms $\propto F^2$ in the power spectrum one should keep terms $\propto F$ and $\propto F^2$ in the response,
\begin{align}
\label{eq:causality}
q(t)\approx &q_0(t)+\int\nolimits_{-\infty}^tdt'\chi_1(t,t')F(t') \nonumber\\
& + \iint\nolimits_{-\infty}^t dt' dt''\chi_2(t,t',t'')F(t')F(t'').
\end{align}
Here $q_0(t)$ is thermal displacement in the absence of driving. 
Equation (\ref{eq:causality}) does not include averaging, $\chi_1$ and $\chi_2$ are the fluctuating linear and nonlinear susceptibilities. The standard linear susceptibility is $\langle \chi_1(t,t')\rangle$, it  is a function of $t-t'$. For a harmonic oscillator, which is the central topic of this paper, $\chi_2 = 0$.

The conventionally measured oscillator power spectrum is $\Phi(\omega)= 2{\rm Re}~\int_0^\infty dt e^{i\omega t}\DblAv{ q(t+t')q(t')}$, where $\DblAv{\cdot}$ indicates statistical averaging and averaging with respect to $t'$ over  the driving period $2\pi/\omega_F$. For weak driving
\begin{equation}
\label{eq:correlator_defined}
\Phi(\omega) \approx \Phi_0(\omega) + \frac{\pi}{2}F^2|\chi(\omega_F)|^2\delta(\omega-\omega_F) +F^2\Phi_F(\omega).
\end{equation}
This spectrum is sketched in Fig.~\ref{fig:sketch}. Function $\Phi_0$ is the power spectrum in the absence of driving, a resonant peak associated with thermal vibrations of the oscillator.
The $\delta$-peak at the driving frequency in Eq.~(\ref{eq:correlator_defined}) and in Fig.~\ref{fig:sketch} describes average forced oscillator vibrations, $\chi(\omega)$ is the Fourier transform of $\langle \chi_1(t,t')\rangle$ over $t-t'$. 

Of primary interest to us is the term $\Phi_F(\omega)$, shown by the envelope of the green area in Fig.~\ref{fig:sketch}.  It describes the interplay of frequency fluctuations and the driving. We consider it for $\omega$ close to $\omega_F$ assuming a high quality factor, $\omega_0 t_r\gg 1$, typical for mesoscopic systems, and resonant driving, $|\omega_F-\omega_0|\ll \omega_F$.

For a harmonic oscillator with fluctuating frequency, $\omega_{\rm osc}(t)=\omega_0+\xi(t)$, where $\xi(t)$ is zero-mean noise. We assume that the noise is weak compared to $\omega_0$ and that its correlation time $t_c\gg \omega_0^{-1}$. The noise then does not cause parametric excitation of the oscillator \cite{Lindenberg1981,Gitterman_book2005}.

The most simple model of the oscillator dynamics is described by equation $\ddot q + 2\Gamma \dot q + [\omega_0^2 + 2\omega_0\xi(t)]q = F\cos\omega_F t + f(t)$, where $f(t)$ is thermal noise and $\Gamma=t_r^{-1}$ is the relaxation rate. Both $f(t)$ and the direct frequency noise $\xi(t)$ lead to fluctuations of the oscillator phase. Separating their contributions by measuring the commonly used Allan variance (cf. \cite{Cleland2002}) is complicated. However, these two types of noise have different physical origin, and our results show how they can be separated using the power spectrum; a different approach, which however may not be implemented with a standard spectrum analyzer, was proposed in \cite{Maizelis2011}.

In the standard rotating wave approximation, see Appendix, the fluctuating linear susceptibility of a damped harmonic oscillator is,
\begin{align}
\label{eq:harmonic_chi_1}
&\chi_1(t,t') =  \frac{i}{2\omega_0}
e^{-(\Gamma+i\omega_0) (t-t')-i\int\nolimits_{t'}^tdt'' \xi(t'') } + {\rm c.c.}
\end{align}
($\chi_2=0$). Equation~(\ref{eq:harmonic_chi_1}) often applies even where the oscillator dynamics in the lab frame is non-Markovian.

Explicit expressions for $\chi_1$ and $\Phi_F(\omega)$ can be obtained from Eq.~(\ref{eq:harmonic_chi_1}) in the limiting cases.  For weak frequency noise, one can expand $\chi_1$ in $\xi(t)$. To the leading order, the spectrum $\Phi_F$ is proportional to the noise power spectrum $\Xi(\Omega)=\int_{-\infty}^\infty  dt \Av{\xi(t)\xi(0)}\exp(i\Omega t)$,
\begin{align}
\label{eq:weak_noise}
&\Phi_F(\omega)\approx \frac{ 1}{16\omega_0^2[\Gamma^2+(\omega_F-\omega_0)^2]}\frac{\Xi(\omega - \omega_F)}{\Gamma^2+(\omega-\omega_0)^2}.
\end{align}
This expression provides a direct means for measuring the frequency noise spectrum. It already shows the peculiar features qualitatively discussed above. If $\Xi(\Omega)$ peaks at zero frequency and is narrow on the scale $\Gamma$ (as for $1/f$-type noise, for example),  $\Phi_F(\omega)$ has a peak at $\omega_F$, cf. Fig.~\ref{fig:sketch}a. The shape of this peak coincides with that of $\Xi(\Omega)$. If, on the other hand, $\Xi(\Omega)$ is almost flat on the frequency scale $\Gamma,|\omega_F-\omega_0|$ (broad-band noise), $\Phi_F(\omega)$ has a Lorentzian peak at $\omega_0$, cf. Fig.~\ref{fig:sketch}b. 

To describe the effect of a narrow-band, but not necessarily weak frequency noise, one can replace  $\xi(t'')$ in Eq.~(\ref{eq:harmonic_chi_1}) with $\xi(t)$. This corresponds to the ``instantaneous" slowly fluctuating susceptibility $i/2\omega_0[\Gamma-i(\omega_F-\omega_0-\xi(t))]$. The resulting narrow spectrum $\Phi_F(\omega)$ is determined by the spectrum and statistics of the frequency noise. The simple relation (\ref{eq:weak_noise}) between $\Phi_F(\omega)$ and $\Xi(\omega)$ follows from this analysis for $\langle\xi^2\rangle \ll \Gamma^2+(\omega_F-\omega_0)^2$. Importantly, this condition can be achieved by tuning $\omega_F$ somewhat away from $\omega_0$.

The case of flat $\Xi(\Omega)$, i.e., of $\xi(t)$ being $\delta$-correlated on time scale $t_r$, can be analyzed for an arbitrary  noise strength using that the characteristic functional of a $\delta$-correlated noise is ${\cal P}[k(t)]=\langle\exp[i\int dt k(t)\xi(t)]\rangle = \exp[-\int \mu(k(t)) dt]$, where function $\mu(k)$ is determined by the noise statistics.
Then from Eq.~(\ref{eq:harmonic_chi_1})
\begin{align}
\label{eq:white_noise}
\Phi_F(\omega)=\frac{[\Re\, \mu(1)]/\Gamma}{8\omega_0^2[\tilde\Gamma ^2 +(\omega_F -\tilde\omega_0)^2]}
\frac{\tilde\Gamma}{\tilde\Gamma^2+ (\omega-\tilde\omega_0)^2}.
\end{align}
The spectrum (\ref{eq:white_noise}) and the spectrum $\Phi_0(\omega)$ in the absence of periodic driving have the same shape given by the last factor in (\ref{eq:white_noise}): a Lorentzian centered at the noise-renormalized oscillator eigenfrequency $\tilde\omega_0=\omega_0-\Im\,\mu(1)$ with halfwidth $\tilde\Gamma = \Gamma + \Re\,\mu(1)$. However, in contrast to $\Phi_0(\omega)$, the area of $F^2\Phi_F(\omega)$ is independent of the intensity ($\propto k_BT$) of the dissipation-related noise. Instead it is proportional to the frequency-noise characteristic $\Re\,\mu(1)$. Equation (\ref{eq:white_noise}) suggests how to separate the noise-induced broadening of the oscillator spectrum from the decay-induced broadening, see below.

\begin{figure}[h]
\includegraphics{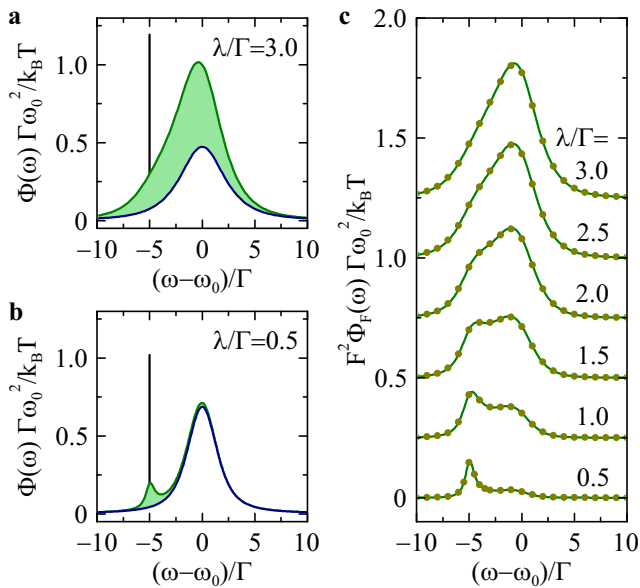} 
\caption{The power spectrum of the oscillator with a Gaussian frequency noise with the spectrum $\Xi(\Omega)=2D\lambda^2/(\lambda^2+\Omega^2)$. The noise intensity is $D/\Gamma =2$. Panels a and b: the full spectrum. The color coding is the same as in Fig.~\ref{fig:sketch}, $F^2/16\Gamma^2=20 k_BT$. Panel c: the driving-induced term. The solid lines and dots show the analytic theory and simulations;  the consecutive curves are shifted by 0.25 along the ordinate.}
\label{fig:theory_1}
\end{figure}

Function $\Phi_F$ can be found in a closed form for a Gaussian noise $\xi(t)$. The results are shown in Fig.~\ref{fig:theory_1} for the noise power spectrum with bandwidth $\lambda$,  $\Xi(\Omega) = 2D\lambda^2/(\lambda^2 + \Omega^2)$. They illustrate how the shape of $\Phi_F(\omega)$ changes from a peak at $\omega_F$ for a narrow-band noise ($\lambda \ll \Gamma$) to a peak at $\omega_0$ for a broadband noise ($\lambda\gg \Gamma$). The overall area of the spectrum $\Phi_F$ nonmonotonically depends on the frequency noise intensity: it is linear in the noise intensity for weak noise, cf. Eq.~(\ref{eq:weak_noise}), but for a large noise intensity it decreases, since the decoherence rate of the oscillator increases.

To corroborate the theory, we measured the spectrum of a modulated carbon nanotube resonator at $T=$~1.2K. For such $T$ and weak driving, low-lying flexural modes of the resonator are well described by harmonic oscillators. The driving was applied as an ac voltage $\delta V_g$ on the gate electrode, and the power spectrum of the mechanical vibrations was probed by measuring the noise in the current flowing through the nanotube \cite{Moser2013}. 

\begin{figure}[ht]
\includegraphics{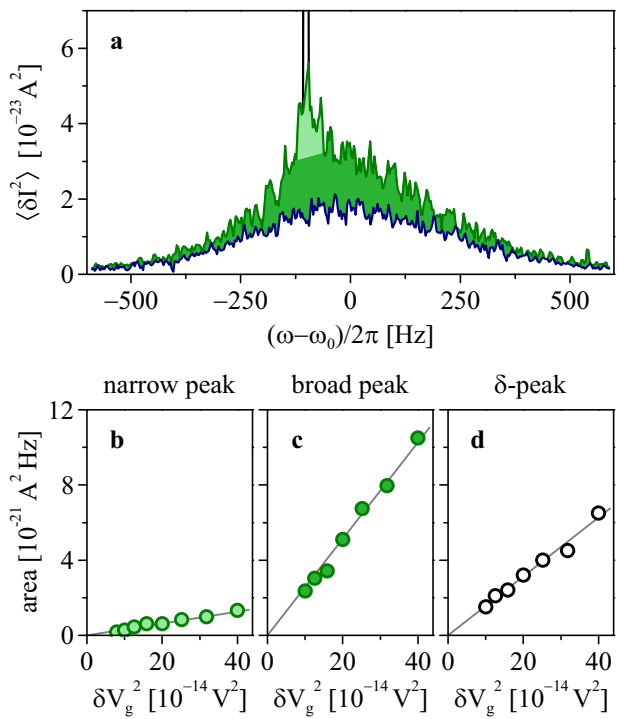} 
\caption{(a) The power spectrum of the fluctuating current $\delta I(t)$ through a driven carbon nanotube. The measurement bandwidth is 4.7~Hz. The eigenfrequency of the studied flexural mode is 6.3~MHz. The driving frequency is 100~Hz below the resonance frequency. The blue line refers to the power spectrum without driving; the green area shows the driving-induced spectral change. This change is separated into the broad peak (darker green), narrow peak (lighter green), and a delta-spike at the modulation frequency. This spike lies within 3 bins, within our experimental resolution, and is represented by the black vertical lines. The separation of the broad and narrow peaks is done by the straight line that interpolates the broad peak. Shown in the lower panels is the dependence of the lighter green area (b), the darker green area (c), and the area under the $\delta$-peak (d) on the squared amplitude of the modulating gate voltage; as expected from the theory, it is close to linear.}
\label{fig:exper}
\end{figure}

Figure 3a compares the power spectra of the nanotube vibrations obtained with and without the oscillating force. The spectrum without modulation (blue trace) is close to a Lorentzian, as expected for thermal vibrations. The spectrum with modulation (green trace)  displays a narrow peak centered at the modulation frequency and a much broader peak of the same shape as the spectrum without modulation. The areas of the both modulation-induced parts of the spectrum scale as $\delta V_g^2$ (Fig.~\ref{fig:exper}b, c), in agreement with Eq.~(2). The separation between the parts is subject to some uncertainty because of the measurement noise in Fig.~\ref{fig:exper}a, see Appendix. The resulting uncertainty in the slopes in Fig.~\ref{fig:exper}b, c is $\lesssim 10$\%. The change of the spectrum is not a heating effect associated with the modulation, since we estimate temperature increases as $\lesssim 10^{-8}$~K. The spectral feature at $\omega_F$ is not related to the phase noise of the source used for the modulation; indeed, the phase noise of our source $\approx 10$~Hz away from $\omega_F$ could only account for $\approx 0.01$~\% of the measured power spectrum.

The driving-induced spectral change provides a simple means for estimating the intrinsic relaxation rate of the resonator $\Gamma$ from our experimental data. This can be done using the areas  $F^2S_{\rm nb}$ and  $F^2S_{\rm bb}$ of the narrow and broad peaks in Fig.~\ref{fig:exper}, respectively.  Comparing them to the area under the driving-induced $\delta$-peak $F^2S_\delta$, we eliminate $F$ and obtain from Eq.~(\ref{eq:white_noise})
\begin{equation}
\tilde{\Gamma}/\Gamma \approx 1+(S_{\rm bb}/S_{\delta})\left[1 - (S_{\rm nb}/S_{\delta})\right];
\end{equation}
we used $S_{\rm nb} \ll S_\delta$. With $\tilde{\Gamma}/(2\pi)\simeq 230$~Hz read out from the collected spectra (such as the one in Fig.~3a), along with $S_{\rm bb}/S_{\delta}$ and $S_{\rm nb}/S_{\delta}$ measured from Figs.~3b-d, we obtain $\tilde{\Gamma}/\Gamma\simeq2.1$, which gives $\Gamma/(2\pi) \simeq110$~Hz. Therefore, the broad-band fluctuations of the resonant frequency account for $\gtrsim 50$\% of the measured mechanical linewidth. Because of the noise in the measurement in Fig.~3a, the uncertainty in $\Gamma$ is $\lesssim 10$\%. 

The narrow-band frequency noise can also be characterized from the measurements. Its power spectrum is $\propto 1/|\omega-\omega_F|^\alpha$ with $\alpha \approx 1/2$, see Appendix. Obtaining the power spectra $\Phi_F$ for several values of $\omega_F-\omega_0$ should allow separating the low-frequency part of the frequency noise spectrum $\Xi(\omega)$ even where it is not weaker than the broad-band part, and reading it directly off the data on the power spectrum using Eq.~(\ref{eq:weak_noise}).

An important contribution to spectral broadening of mesoscopic oscillators can come from their nonlinearity \cite{Dykman2012b}. Because the vibration frequency of a nonlinear oscillator depends on the vibration amplitude, 
thermal fluctuations of the amplitude lead to frequency fluctuations.  This makes the power spectrum non-Lorentzian and asymmetric even in the absence of driving \cite{DK_review84}. The shape of the spectrum is determined by the interrelation between the frequency uncertainty $\Gamma$ due to the oscillator decay and the width $\Delta\omega$ of the frequency distribution due to thermal distribution of the vibration amplitude. In the presence of driving, superimposed on this effect is the frequency shift due to the driving-induced vibrations.

\begin{figure}[]
\includegraphics
{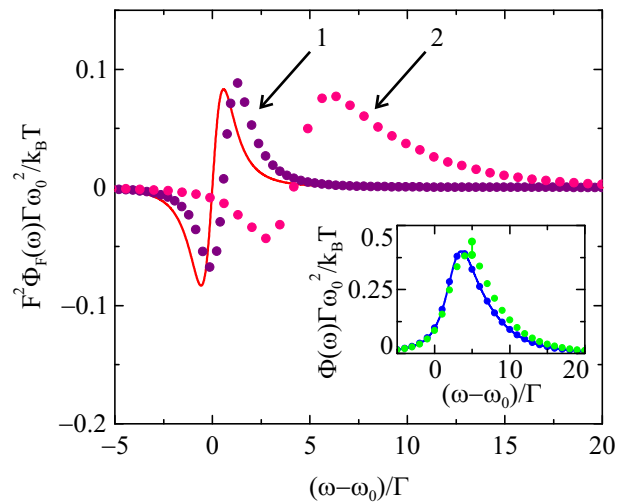}
\caption{The driving-induced part of the spectrum of a nonlinear oscillator.  The dots show the results of simulations. The red line shows the analytical results for $\Phi_F(\omega)$ for small $\Delta\omega/\Gamma$ for the parameters of curve 1. The scaled values of the nonlinearity parameter, the detuning, and the driving strength on the curves  1 and 2 are, respectively, $\Delta\omega/\Gamma =$~0.125 and 1.25, $(\omega_F-\omega_0)/\Gamma =$~0.5 and 5, and $3\gamma F^2/32\omega_0^3(\omega_F-\omega_0)^3 =$~0.64 and 0.01. The inset shows the full spectrum  for the parameters of curve 2 (green dots, simulations); the spectrum without driving for the same $\Delta\omega/\Gamma$ is shown by the solid line (analytical) and blue dots on top of it (simulations).}
\label{fig:nonlinear_osc}
\end{figure}

The strong dependence of the driving-induced spectral change on the ratio $\Delta\omega/\Gamma$ is illustrated in Fig.~\ref{fig:nonlinear_osc} for the important model where the nonlinear term in the oscillator energy is $\gamma q^4/4$, so that $\Delta\omega \approx 3|\gamma|k_BT/8\omega_0^3$. For small $\Delta\omega/\Gamma$ the major effect of the driving is the shift of the spectrum; then $\Phi_F(\omega)\propto \partial_\omega\Phi_0(\omega)\propto (\omega-\omega_0)\Phi_0(\omega)$ has a characteristic dispersive shape, changing sign at $\omega_0$. With increasing $\Delta\omega/\Gamma$ the shape of $\Phi_F(\omega)$ becomes more complicated. Generally it still has positive and negative parts, in dramatic difference from the case of a harmonic oscillator with fluctuating frequency. Also, in contrast to a harmonic oscillator, keeping terms $\propto F^2$ in the power spectrum of a nonlinear oscillator is justified only for weak modulating fields. A detailed theory of $\Phi_F$ for nonlinear oscillators based on Eq.~(\ref{eq:causality}) will be presented elsewhere.

The above results show that the interplay of driving and  frequency noise qualitatively changes oscillator spectra compared to the spectra with no frequency fluctuations \cite{Heitler2010}. The change sensitively depends on the frequency noise intensity and power spectrum. The possibility to separate contributions from different parts of the frequency-noise spectrum is of significant interest, as they may come from physically different sources, like two-level fluctuators and dispersive coupling to other modes, to mention but a few. 

The results suggest a way of discriminating between three major factors of the broadening of the oscillator spectra: decay (energy relaxation), frequency fluctuations induced directly by the noise that modulates the eigenfrequency, and frequency fluctuations due to the oscillator nonlinearity.  For linear oscillators, our simple procedure yields the decay rate without the need of an actual ring-down measurement that is often difficult to implement, in particular for nanotube mechanical resonators. We also find that the nonlinearity-induced spectral change has a qualitatively different shape from that due to frequency noise in linear oscillators.

The analysis of driven linear oscillators with a fluctuating frequency immediately extends to the quantum regime, which is attracting much interest in nano- and optomechanics  \cite{Aspelmeyer2013,Blencowe2004a,Schwab2005a,Clerk2010a,O'Connell2010,Meenehan2014}. For nonlinear oscillators, the nonequidistance of the energy levels can bring in new features compared to the classical limit.    

This research was supported in part by the US Army Research Office (W911NF-12-1-0235), US Defense Advanced Research Agency (FA8650-13-1-7301), the ERC-carbonNEMS project (279278), the grant MAT2012-31338 from the Spanish state, and the grant AGAUR-SGR of the Catalan government.

\appendix
\section{General expression for the power spectrum}

The explicit expression for the driving-induced term in the power spectrum of fluctuations of the oscillator reads
\begin{align}
\label{eq:Phi_F}
&\Phi_F(\omega)=\frac{1}{2}\Re \int_0^{\infty} dt e^{i(\omega-\omega_F)t}\iint_{-\infty}^0 d\tau d\tau'e^{i\omega_F(\tau' - \tau)} \nonumber\\
&\times\Av{\chi_1(t,t+\tau)\left[\chi_1(0,\tau')-\Av{\chi_1(0,\tau')}\right]}+ \Phi_F^{(2)}(\omega).
\end{align} 
This expression follows from Eqs.~(1) and (2) of the main text. The first term gives the contribution of the fluctuations of the linear susceptibility. The second term gives the contribution from the nonlinear susceptibility,
\begin{align}
\label{eq:Phi_2}
&\Phi_F^{(2)}(\omega)=\Re\int_0^{\infty} dt e^{i\omega t}\iint_{-\infty}^0 d\tau d\tau' 
\cos[\omega_F(\tau - \tau')]\nonumber\\
&\times \left[\Av{\chi_2(t,t+\tau,t+\tau') q_0(0)} + \Av{q_0(t)\chi_2(0,\tau,\tau')}\right].
\end{align}
This term describes the correlation between fluctuations of the second-order susceptibility and thermal fluctuations in the absence of periodic driving. We emphasize that, for a resonantly modulated underdamped oscillator, it is pronounced at frequencies $\omega$ close to the driving frequency $\omega_F$, not $2\omega_F$. 
Equation (\ref{eq:Phi_2}) describes, in particular, the contribution to the spectrum from the nonlinear susceptibility of a nonlinear oscillator. It is especially convenient in the case of weak nonlinearity, where the oscillator spectrum $\Phi_0(\omega)$ is broadened primarily by the decay rather than by frequency fluctuations due to the interplay of the nonlinearity and the  amplitude fluctuations. In this case the term $\Phi_F^{(2)}$ gives the main contribution to $\Phi_F$. The theory of a nonlinear oscillator will be discussed in a separate publication.


\section{Averaging over frequency fluctuations for a linear oscillator}

Equation~(3) of the main text for the susceptibility of a linear underdamped oscillator with fluctuating frequency can be found in a standard way by changing from the fast oscillating variables $q,\dot q$ to slow complex oscillator amplitude $u(t)=[q(t)+(i\omega_F)^{-1}\dot q(t)]\exp(-i\omega_Ft)/2$. If the equation of motion in the lab frame is Markovian, $\ddot q + 2\Gamma \dot q + [\omega_0^2 + 2\omega_0\xi(t))]q = F\cos\omega_F t + f(t)$, where $f(t)$ is the dissipation-related thermal noise, as in the example discussed in the main text, the equation for $u(t)$ in the rotating wave approximation reads
\begin{equation}
\label{eq:dot_u}
\dot u =  -[\Gamma + i\delta\omega_F  - i\xi(t)]u -i \frac{F}{4\omega_0} + f_{u}(t).
\end{equation}
Here, $\delta\omega_F = \omega_F-\omega_0$ is the detuning of the driving frequency from the oscillator eigenfrequency; $f_{u}(t)=[f(t)/2i\omega_0]\exp(-i\omega_0t)$. Equation~(\ref{eq:dot_u}) applies on the time scale that largely exceeds $ \omega_0^{-1}$. On this scale $f_{u}(t)$ is $\delta$-correlated even where in the lab frame the oscillator dynamics is non-Markovian, cf \cite{DK_review84}. Solving the linear equation (\ref{eq:dot_u}), one immediately obtains  Eq.~(3) of the main text for the oscillator susceptibility $\chi_1(t,t')$. We disregard corrections $\sim |\delta\omega_F| /\omega_F$; in particular in Eq.~(\ref{eq:dot_u}) for convenience we replaced $F/\omega_F$ with $F/\omega_0$; similarly, in the expression for $f_u$ we replaced $f/\omega_F$ with $f/\omega_0$. 

We note that the noise $f_u(t)$ drops out from the moments $\langle u^n(t)\rangle$ \cite{Maizelis2011}. This can be used to characterize the statistics of the frequency noise. In this paper we consider the change of the conventionally measured characteristic, the power spectrum, and the extra spectral features related to the interplay of the driving and frequency noise.

It is convenient to rewrite Eq.~(\ref{eq:Phi_F}) for the spectrum $\Phi_F(\omega)$ near its maximum in the form that explicitly takes into account that, when the expression for the susceptibility is substituted into Eq.~(\ref{eq:Phi_F}), the fast-oscillating terms in the integrands can be disregarded. This gives 
\begin{align}
\label{eq:Phi_F_convenient}
&\Phi_F(\omega)=(8\omega_0^2)^{-1}\Re \int_0^\infty dt \exp[i(\omega-\omega_F)t]\nonumber\\
&\times \int_{-\infty}^t dt'\int_{-\infty}^0 dt_1'\bigl\langle\chi_{\rm sl}(t,t')[\chi_{\rm sl}^*(0,t_1')-\Av{\chi_{\rm sl}^*(0,t_1')}]\bigr\rangle,\nonumber\\
&\chi_{\rm sl}(t,t')=e^{-(\Gamma - i\delta\omega_F)(t-t')} \exp\left[-i\int\nolimits_{t'}^tdt'' \xi(t'')\right].
\end{align}
Here, function $\chi_{\rm sl}(t,t')$ gives the slowly varying factor in the fast-oscillating time-dependent oscillator susceptibility $\chi_1(t,t')$. Function $\Av{\chi_{\rm sl}(0,t)}\equiv \Av{\chi_{\rm sl}(-t,0)}$ gives the standard (average) susceptibility 
\begin{align}
\label{eq:LRT}
&\chi(\omega_F)=\int_0^\infty dte^{i\omega_F t}\Av{\chi_1(t,0)} = \frac{i}{2\omega_0}\int_0^\infty dt \Av{\chi_{\rm sl}(t,0)}.
\end{align}
The mean forced displacement of the oscillator in the linear response theory is $\Av{q(t)}=\frac{1}{2}Fe^{-i\omega_F t}\chi(\omega_F) + {\rm c.c.}$. 

\subsection{Noise averaging for fast, slow, and Gaussian noise}

Averaging over  $\xi(t)$ in Eqs.~(\ref{eq:Phi_F_convenient}) and (\ref{eq:LRT}) can be done using the noise characteristic functional (cf. \cite{FeynmanQM}),
\[{\cal P}[k(t)]=\left\langle\exp\left[i\int dt\,k(t)\xi(t)\right] \right\rangle.\]
As seen from Eq.~(\ref{eq:Phi_F_convenient}), function $\Av{\chi_{\rm sl}(t,t')}$ is determined by ${\cal P}[k(t'')]$ with $k(t'') =-1$ if $t'<t''<t$ and $k(t)=0$ otherwise. For $\delta$-correlated noise, where ${\cal P}[k(t)]=\exp[-\int dt\,\mu(k(t))]$, taking into account that $\mu(0) =(d\mu/dk)_{k=0} =0$ and $\mu(-k)=\mu^*(k)$, we obtain
\begin{align}
\label{eq:susceptibility_delta_correlated}
&\Av{\chi_{\rm sl}(t,t')} = \exp\bigl[- \left(\Gamma -i\delta\omega_F + \mu^*(1)] \right)(t-t')\bigr], \nonumber\\
&\chi(\omega_F) = (i/2\omega_0)\left[\tilde\Gamma - i(\omega_F-\tilde\omega_0)\right]^{-1}
\end{align}
with $\tilde\Gamma = \Gamma+ {\rm Re}\,\mu(1)$ and $\tilde\omega_0=\omega_0 - {\rm Im}\,\mu(1)$. Thus, frequency noise leads to the broadening of the conventional susceptibility $\Re\, \mu(1)$ and the effective shift of the oscillator eigenfrequency by $-\Im\, \mu(1)$. We note that the noise can be considered $\delta$-correlated when its spectrum is flat not just  on the scale $\gtrsim \Gamma$, but on the scale $\gtrsim \Gamma + \Re~\mu(1)$, which itself depends on the noise intensity. At the same time, the noise spectrum is assumed to be much narrower than $\omega_0$. As seen from Eq.~(\ref{eq:Phi_F_convenient}) the noise components oscillating at frequencies much higher than $\Gamma + {\rm Re}\,\mu(1), |\delta\omega_F|$ are averaged out; frequency noise with frequencies $\sim\omega_0$ was disregarded in Eq.~(\ref{eq:dot_u}). When writing Eq.~(\ref{eq:dot_u}) we also assumed that noise at frequencies close to $2\omega_0\approx 2\omega_F$ is very weak and can be disregarded. If this were not the  case, one would have to take into account the effects of nonlinear friction that come from the coupling to the source of the noise, cf. \cite{DK_review84}. 

Averaging  the term $\Av{\chi_{\rm sl}(t,t')\chi^*_{\rm sl}(0,t_1')}$ in Eq.~(\ref{eq:Phi_F_convenient}) comes to calculating 
\begin{align}
\label{eq:k(t)_definition}
&\left\langle\exp\left[-i\int_{t'}^t dt''\xi(t'') + i\int_{t_1'}^0 dt_1''\xi(t_1'')\right]\right\rangle \nonumber\\
&\equiv\left\langle\exp\left[i\int_{-\infty}^{\infty} dt_2k(t_2)\xi(t_2)\right]\right\rangle.
\end{align}
Here $t>0$ and $-\infty < t' \leq t, -\infty < t_1'\leq 0$. Clearly, in this equation $k(t_2)=0,\pm 1$.
For $t'<0$ we have $k(t_2)=\sgn(t'-t'_1)$ if $\min(t',t'_1)<t_2 <\max(t_1,t'_1)$ and $k(t_2)=-1$ if $0<t_2<t$; for $t'>0$ we have $k(t_2)=1$, if $t_1'<t_2<0$ and $k(t_2)=-1$, if $t'<t_2<t$; otherwise $k(t_2)=0$.
For a $\delta$-correlated noise the averaging using the explicit form of ${\cal P}[k(t)]$ and integration over $t',t_1',t$ gives Eq.~(5) of the main text.

For a stationary Gaussian noise the characteristic functional is expressed in terms of the noise correlator \cite{FeynmanQM},
\[{\cal P}[k(t)]=\exp\left[-\frac{1}{2}\int dt\,dt'\,\langle \xi(t)\xi(t')\rangle k(t)k(t')\right].\] 
If the correlator $\langle \xi(t)\xi(t')\rangle$ or equivalently, the power spectrum $\Xi(\Omega)$, are known, using the values of $k(t)$ given below Eq.~(\ref{eq:k(t)_definition}) one can perform the averaging in Eq.~(\ref{eq:Phi_F_convenient}) and then perform integration over time to find the power spectrum $\Phi_F$. This was done to obtain the results shown in Fig.~2 of the main text.

For slowly varying frequency noise on the scale of the oscillator relaxation time $\Gamma^{-1}$, the evaluation of the susceptibility following the prescription given in the main text leads to expression
\begin{equation}
\label{eq:susceptibility_slow}
\chi(\omega_F) = \frac{i}{2\omega_0}\langle X(t)\rangle,\quad X(t)=\left[\Gamma-i\delta\omega_F+i\xi(t)\right]^{-1}.
\end{equation}
whereas the expression for the driving-induced term in the power spectrum  reads
\begin{align}
\label{eq:slow_noise}
&\Phi_F(\omega) \approx \frac{1}{8\omega_0^2}\Re\int_0^\infty dt e^{i(\omega-\omega_F)t}
\bigl\langle X(t) [X^*(0)\nonumber\\
&-\Av{X^*(0)}]\bigr\rangle.
\end{align}
These expressions can be used for numerical calculations if the statistics of the noise $\xi(t)$ is known.

\subsection{The weak-noise condition}

In the limit of weak slow noise, $\langle \xi^2(t)\rangle \ll |\Gamma -i\delta\omega_F|^2$, Eq.~(\ref{eq:slow_noise}) goes over into the result for such noise obtained in the main text; note that in Eq.~(4) of the main text one should replace $\omega-\omega_0$ with $\omega_F - \omega_0$ in the slow-noise limit, since function $\Xi(\Omega)$ is concentrated in the range of small $\Omega\ll \Gamma$.  For the broad-band noise, on the other hand, the weak-noise limit discussed in the main text corresponds to $|\mu(1)|\ll \Gamma$. In this case the noise power spectrum is flat and $\Xi(\Omega)=(d^2\mu/dk^2)_{k=0}\sim|\mu(1)|\ll \Gamma$. Generally, the weak noise condition used to obtain Eq.~(4) of the main text certainly holds for $\max\Xi(\Omega) \ll \Gamma$. It is important that, for slow noise, the condition is less stringent and can be met by increasing the detuning $|\delta\omega_F|$, allowing one to read the slow-noise power spectrum directly off the oscillator power spectrum.

\subsection{Susceptibility of a linear oscillator with weakly fluctuating frequency}

Both the standard susceptibility $\chi(\omega)$ and the power spectrum in the absence of driving $\Phi_0(\omega)$ are affected by frequency noise. In the considered case they are related by the fluctuation-dissipation relation, $\Phi_0(\omega)= (2k_BT/\omega){\rm Im}~\chi(\omega)$. For a non-white frequency noise the spectrum $\Phi_0(\omega)$ becomes non-Lorentzian. 

The explicit expressions for the susceptibility in the limiting cases of fast and slow frequency noise were given above, Eqs.~(\ref{eq:susceptibility_delta_correlated}) and (\ref{eq:susceptibility_slow}). A simple explicit expression for $\chi(\omega)$  follows from Eqs.~(\ref{eq:Phi_F_convenient}) and (\ref{eq:LRT}) also in the case of weak noise. Here, the susceptibility becomes
\begin{align}
\label{eq:correction_to_chi}
\chi(\omega)\approx \frac{i}{2\omega_0(\Gamma-i\delta\omega)}\left[1-\int\frac{d\Omega}{2\pi(\Gamma-i\delta\omega)}\frac{\Xi(\Omega)}
{\Gamma-i\delta\omega-i\Omega}\right],
\end{align}
where $\Xi(\Omega)$ is the frequency noise power spectrum and $\delta\omega = \omega-\omega_0$. Importantly,  the noise-induced correction just slightly distorts the susceptibility. For example, a sharp low-frequency peak of $\Xi(\Omega)$ does not lead to a narrow peak in $\chi(\omega)$ and, respectively, in the power spectrum $\Phi_0(\omega)$. This should be contrasted with the narrow peak in $\Phi_F(\omega)$, which emerges in this case.

\hfill

\section{Power-law noise in carbon nanotube resonators}

\begin{figure}[h]
\includegraphics{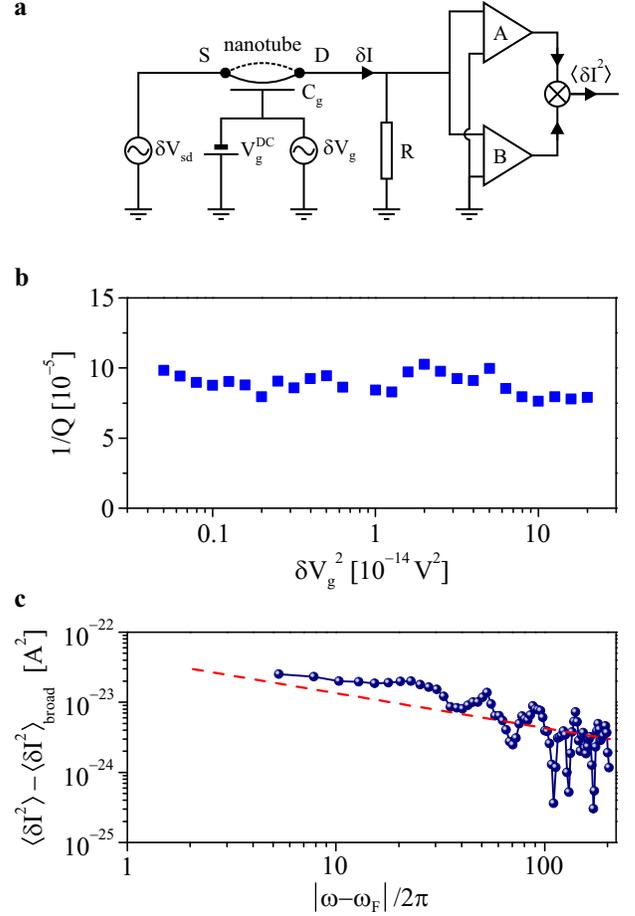}
\caption{(a) Measurement setup. Fluctuations of the position of
the nanotube induce fluctuations of the gate capacitance $C_{g}$,
which in turn result in fluctuations of the conductance of the
nanotube. Applying an oscillating voltage $\delta V_{sd}(t)$
between source S and drain D results in current fluctuations
$\delta I$, which are converted into voltage fluctuations across a
resistor $R$. These voltage fluctuations are amplified and
cross-correlated, yielding the variance of current fluctuations
$\langle\delta I^{2}\rangle$. Modulation at frequency $\omega_{F}$
is obtained by applying an oscillating gate voltage of amplitude
$\delta V_{g}$. (b) Inverse of the quality factor as a function of
$\delta V_{g}^{2}$. Parameters used are $V_{g}^{DC}=1.454$~V,
$\delta V_{sd}=89\times10^{-6}$~V, and
integration time 32~s. The resonant frequency is
$\omega_{0}/(2\pi)=6.3\times10^{6}$~Hz. (c) Narrow band
frequency noise spectrum. It is
obtained by fitting the broad band frequency noise part
$\langle\delta I^{2}\rangle_\textrm{broad}(\omega)$ of the
experimental spectrum in Fig.~3 to a Lorentzian, and then by
subtracting this fit from the experimental spectrum. The red line
is a fit to $1/f^{1/2}$, where
$f=|\omega-\omega_{F}|/2\pi$.}\label{Figexp}
\end{figure}

The device consists of a carbon nanotube contacted by source and
drain electrodes and suspended over a gate electrode. Details of
the fabrication and the geometry of the device can be found in
Ref.~\cite{Moser2013}. We measure power spectra of
displacement fluctuations using the experimental setup sketched in
Fig.~\ref{Figexp}a. Displacement fluctuations induce conductance
fluctuations. We parametrically down-convert these conductance fluctuations by
applying an AC voltage $\delta V_{sd}(t)$ between source and drain 
at a non-resonant frequency $\omega_{sd}$, resulting in current
fluctuations at frequencies $|\omega_{0}-\omega_{sd}|\sim
2\pi\times 10$~kHz.\\

The spectrum shown in Fig.~3a of the main text is obtained in the
presence of a near resonant oscillating electrostatic force
$\delta F(t)$. This force is created by applying an oscillating
voltage $\delta V_{g}(t)=\delta V_{g}^{AC}\cos\omega_{F}t$ at a
frequency $\omega_{F}=\omega_{0}-2\pi\times102$~Hz, with
$\omega_{0}/(2\pi)=6.3\times10^{6}$~Hz, and an amplitude $\delta
V_{g}^{AC}=4.9\times10^{-7}$~V. In this experiment, a DC gate
voltage $V_{g}^{DC}=1.454$~V and an AC source-drain voltage of
amplitude $\delta V_{sd}^{AC}=89\times10^{-6}$~V are used. The
amplitude $\delta V_{sd}^{AC}$ is kept below the threshold beyond
which the variance of displacement of the nanotube increases with
$\delta V_{sd}^{AC}$ (as in Ref.~\cite{Moser2013}). The mode
temperature is 1.2~K. The integration time is
32~s.\\

It is important to verify that applying $\delta V_{g}(t)$ does not result
in an increase in the mode temperature. We consider the case
$\omega_{F}=\omega_{0}$ where an increase of temperature, if any,
should be most pronounced. Two mechanisms are liable to increase
the mode temperature: (i) dissipated power related to the work
done by the oscillating resonant force from the gate electrode $\delta F(t)$, and (ii) Joule
heating related to the current, flowing through the nanotube, that
is induced by the time-varying capacitance between the nanotube
and the gate electrode. We now discuss the effects of these mechanisms.

\begin{enumerate}[(i)]
\item{From the work of a resonator subject to an oscillating force $\delta F$, the time average power reads:
\begin{equation}
\langle P_{\delta F}\rangle=\frac{\delta F^{2}Q}{2M\omega_{0}}\,,
\end{equation}
where $Q$ is the quality factor and $M$ is the effective mass of
the mode. The amplitude of the oscillating force is $\delta
F=C_{g}^{\prime}V_{g}^{DC}\delta V_{g}$, where $C_{g}^{\prime}$ is
the derivative of the gate capacitance with respect to a small
displacement (we assume that the whole length of the nanotube is
at a single, well-defined potential). From Coulomb blockade
measurements, we estimate that
$C_{g}^{\prime}=1.2\times10^{-12}$~F/m as detailed in
Ref.~\cite{Moser2013}. We estimate the mass
$M=9.8\times10^{-21}$~kg from the diameter and the length of the
nanotube. In Figs.~3b, c of the main text, the maximum amplitude
$\delta V_{g}$ is $\sim6.4\times10^{-7}$~V. Using
$Q=1.2\times10^{4}$, $V_{g}^{DC}=1.454$~V, and
$\omega_{0}/(2\pi)=6.3\times10^{6}$~Hz, we find that the maximum
dissipated power is $\langle P_{\delta
F}\rangle_\textrm{max}\simeq2\times10^{-20}$~W. This is a
minuscule power.

Using a thermal conductance of $10^{-12}$~W/K, this dissipated
power translates into a temperature increase $\Delta
T\sim10^{-8}$~K, a truly insignificant increase. This thermal
conductance is inferred from two published measurements at liquid
helium temperature. The thermal conductance for a multi-wall
carbon nanotube with a length of 2.5~$\mu$m and a diameter of
14~nm was measured to be $\sim10^{-10}$~W/K \cite{Kim2001}. The
thermal conductivity of aligned single-wall nanotubes was measured
to be $\sim1$~Wm$^{-1}$K$^{-1}$ \cite{Hone2000}. These two
measurements indicate that the thermal conductance is in the range
$10^{-12}-10^{-11}$~W/K for a nanotube with a diameter of 1~nm and
a length of 2~$\mu$m.}

\item{As the nanotube vibrates, the distance that separates it from the gate electrode is modulated,
and so is the gate capacitance $C_{g}$. The driving of $C_{g}$
results in a current at the driving frequency that flows
through the nanotube. On resonance, this current reads:
\begin{equation}
I_{\delta C}(t)=\omega_{0}V_{g}^{DC}\delta C_{g}\sin\omega_{0}t\,,
\end{equation}
where $\delta C_{g}$ is the driving amplitude of $C_{g}$. Note
that $I_{\delta C}(t)$ also has components proportional to
$C_{g}\delta V_{g}$, but these have amplitudes that are several
orders of magnitude smaller than $\omega_{0}V_{g}^{DC}\delta
C_{g}$. The time average dissipated power related to Joule heating
reads
\begin{equation}
\langle P_{d}\rangle=R_{t}\langle I_{\delta C}(t)^{2}\rangle=R_{t}(V_{g}^{DC}\omega_{0}\delta C_{g})^{2}/2\,,
\end{equation}
where $R_{t}$ is the resistance of the nanotube. We estimate
$\delta C_{g}=C_{g}^{\prime}\delta z_{0}\simeq10^{-21}$~F, using
the resonant displacement $\delta
z_{0}=QC_{g}^{\prime}V_{g}^{DC}\delta
V_{g}/(M\omega_{0}^{2})\simeq0.6\times10^{-9}$~m as an
approximation of the motional amplitude. Hence, the dissipated
power is $\langle P_{d}\rangle_\textrm{max}\simeq10^{-22}$~W. Here
again, the induced temperature increase can be neglected.}
\end{enumerate}

Confirming these estimates, Fig.~\ref{Figexp}b shows that the
inverse of the quality factor $1/Q$ does not vary as $\delta
V_{g}^{2}$ increases. Since an increase in temperature would
result in an increase in $1/Q$, this further indicates that
$\delta V_{g}(t)$
does not affect the mode temperature.\\

The spectral feature at $\omega_{F}$, which we associate to a
narrow band frequency noise, is not related to the phase noise of
the source used to supply $\delta V_{g}(t)$. Indeed, the phase
noise of our source $\sim10$~Hz away from $\omega_{F}$ is
$\sim-60$~dB$_\textrm{c}/$Hz, which would result in side bands of
amplitude $\sim10^{-27}$~A$^{2}$. These side bands would then be 4
orders of magnitude smaller than the spectral feature we associate
with narrow band frequency noise.

Since the narrow-band frequency noise in the nanotube is comparatively weak, one can interprete the results using the weak-noise expression for the spectrum Eq. (4) of the main text. Then the shape of the resonator spectrum gives the shape of the noise power spectrum. As seen from Fig.~\ref{Figexp}c, the spectrum is of $1/f^\alpha$ type. Our data indicate that $\alpha$ is close to 1/2. 

\begin{figure}[h]
\includegraphics{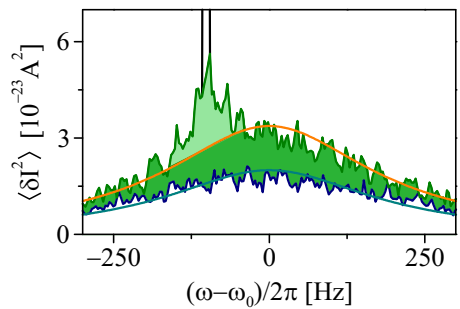}
\caption{The power spectrum of the fluctuating current $\delta I(t)$ shown in Fig. 3a of the main text. Here the separation between the light and dark green shaded areas is obtained by approximating the dark green shaded area by a Lorentzian of the same shape as a Lorentzian that approximates the spectrum without driving (the blue line in Fig.~3a).}
\label{fig:exper_Lorentz}
\end{figure}

To highlight spectral features that we associate to frequency
noise (light and dark green shaded areas in Fig.~3a of the main
text), we exclude the $\delta$-peak at driving frequency
$\omega_{F}$. To this end, we observe that the response of our
signal analyzer to a voltage oscillating at a given frequency is a
delta peak that consists of 3 points above the background.
Similarly, the $\delta$-peak at $\omega_{F}$ displayed as a
black trace in Fig.~3a of the main text consists of 3 data points
above the background signal. We remove those 3 points from the
measured spectra to estimate the spectral areas
plotted in Figs.~3b, c.\\

In Fig.~3a of the main text, the separation between the light and dark green shaded areas is subject to some uncertainty because of the noise in the measurement. In Fig.~\ref{fig:exper_Lorentz} we separate these two areas by approximating the broad peak by a Lorentzian. This also leads to a linear dependence of the areas of the peaks on  $\delta V_g^2$, with the difference in the slopes $\lesssim 10$\% compared to the results in Fig.~3b, c of the main text. Because of the narrow-band noise, the power spectrum without modulation actually differs from a Lorentzian, but this is hard to reveal by measuring just this spectrum alone (or the absorption spectrum).


\section{The area of the driving-induced spectral peak for a linear oscillator}

We now consider the area $S_F$ of the driving induced spectral peak for  $\omega$ close to $\omega_0,\omega_F$; note that this peak may have several maxima, as seen from Fig.~2 of the main text. We define the area as an integral over positive frequencies,  $S_F=\int_0^\infty d\omega\,\Phi_F(\omega)$. Keeping in mind that $\Phi_F(\omega)$ is small for large $|\omega-\omega_F|\sim \omega_F$ [in fact, Eq.~(\ref{eq:Phi_F_convenient}) does not apply for such $\omega$], we obtain
\begin{align}
\label{eq:area}
S_F&=\frac{\pi}{8\omega_0^2}\iint_{-\infty}^0 dt\,dt'\langle\chi_{\rm sl}(0,t)\chi_{\rm sl}^*(0,t')\rangle 
\nonumber\\
&-\frac{\pi}{8\omega_0^2}\left|\int_{-\infty}^0dt\,\langle\chi_{\rm sl}(0,t)\rangle\right|^2.
\end{align}
This expression describes the dependence of the area of the driving-induced spectrum on the parameters and statistics of the frequency noise. 

From Eq.~(\ref{eq:area}), the area $S_F$ becomes zero in the absence of frequency noise, since $\langle \chi_{\rm sl}(t,t')\rangle = \chi_{\rm sl}(t,t')$ in this case. The area $S_F$ linearly increases with the frequency noise intensity for weak noise, as seen from Eq.~(4) of the main text. 

An explicit expression for $S_F$ can be obtained for white frequency noise. From Eq.~(5) of the main text,
\begin{equation}
\label{eq:area_white}
S_F=\frac{\pi}{8\Gamma\omega_0^2}\frac{\Re~\!\mu(1)}{|\Gamma + i(\omega_F - \omega_0)+\mu(1)|^2}.
\end{equation}
From Eq.~(\ref{eq:area_white}), $S_F$ linearly increases with the characteristic noise strength $\Re~\mu(1)$ where it is small, but once the noise becomes strong, $S_F$ decreases with increasing $|\mu(1)|$, with $S_F\propto \Re~\mu(1)/|\mu(1)|^2$ for $|\mu(1)|\gg \Gamma,|\omega_F-\omega_0|$. 

For weak narrow-band frequency noise, from Eq.~(4) of the main text one obtains $S_F$ in terms of the noise variance $ \Av{\xi^2(t)}$ as
\[S_F=\frac{\pi}{8\omega_0^2}\frac{\langle\xi^2(t)\rangle}{[\Gamma^2+(\omega_F-\omega_0)^2]^2}.\]

An explicit expression for $S_F$ can be obtained also for a strong Gaussian noise. We will assume that the noise correlator $\Av{\xi(t)\xi(0)}$ is not fast oscillating and, respectively, the noise spectrum $\Xi(\Omega)$ does not have narrow peaks or dips. For  the noise variance $ \Av{\xi^2(t)}$ much larger than $\Gamma^2,\delta\omega_F^2$, and the squared reciprocal noise correlation time $t_c^{-2}$, from Eq.~(\ref{eq:area})
\begin{equation}
\label{eq:area_strong_noise}
S_F\approx \frac{\pi^2}{8\Gamma\omega_0^2}[2\pi\Av{\xi^2(t)}]^{-1/2}.
\end{equation}

\begin{figure}[ht]
\includegraphics[width = 8truecm]{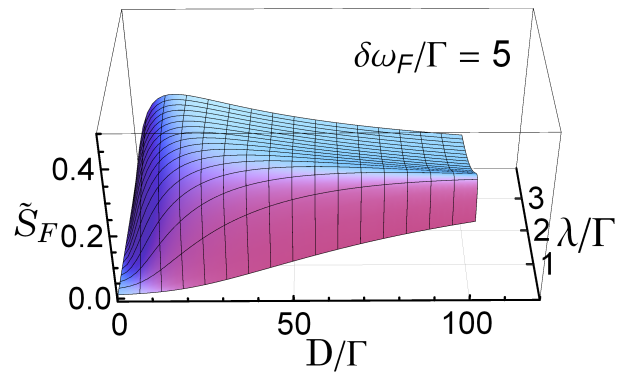}
\caption{The scaled area $\tilde S_F = 8\Gamma^2\omega_0^2S_F$ of the driving-induced peak in the oscillator power spectrum  as a function of the frequency noise parameters.  The data refer to Gaussian frequency noise with the power spectrum $\Xi(\Omega)=2D\lambda^2/(\lambda^2+\Omega^2)$. }
\label{fig:area}
\end{figure}

The variation of $S_F$ with the varying frequency-noise intensity  and bandwidth is shown in Fig.~\ref{fig:area}, which refers to the exponentially correlated Gaussian noise. As seen from this figure, $S_F$ displays a maximum as a function of the noise intensity $D$. The dependence on the noise bandwidth $\lambda$ is more complicated; $S_F$ can have two maxima as a function of $\lambda$ for sufficiently strong noise intensity.

\subsection{Scaling of the driving-induced power spectrum}

A convenient scaling factor for the distribution $\Phi_F(\omega)$ and for the area $S_F$ is provided by the area $S_\delta$ of the $\delta$-peak in the oscillator power spectrum at the driving frequency. As seen from Eq.~(2) of the main text, $S_\delta = (\pi/2)|\chi(\omega_F)|^2$. If $\chi(\omega_F)$ is known from the measured power spectrum in the absence of driving with the invoked fluctuation-dissipation theorem, scaling by $S_\delta$ allows one to avoid the actual measurement of the force $F$, which requires knowledge of the coupling to the driving field. 

The expression for $S_\delta$ simplifies if the frequency noise can be thought of as a sum of a weak narrow-band noise $\xi_{\rm nb}(t)$  and a broad-band ($\delta$-correlated in slow time) noise $\xi_{\rm bb}(t)$, which is not weak, generally, and is statistically independent from the narrrow-band noise. In this case, combining Eqs.~(\ref{eq:susceptibility_delta_correlated}) and (\ref{eq:susceptibility_slow}) and expanding to the leading order in the weak narrow-band noise, we obtain
\begin{align}
\label{eq:area_delta_peak}
S_\delta &\approx \frac{\pi}{8\omega_0^2}[\tilde\Gamma^2+(\omega_F-\tilde\omega_0)^2]^{-1}\nonumber\\
&\times\left[1-\frac{\tilde\Gamma^2 - (\omega_F-\tilde\omega_0)^2}{\pi[\tilde\Gamma^2+(\omega_F-\tilde\omega_0)^2]^2}\int d\omega \,\Xi_{\rm nb}(\omega)\right].
\end{align}
Here, $\Xi_{\rm nb}(\omega)$ is the power spectrum of the narrow-band noise; the variance of the narrow-band noise is  $\langle \xi^2_{\rm nb}(t)\rangle = (2\pi)^{-1}\int d\omega \,\Xi_{\rm nb}(\omega)$.

The correction that contains $\Xi_{\rm nb}$ can be directly read off the area of the narrow peak  $\Phi_F^{\rm (nb)}(\omega)$  in the spectrum $\Phi_F(\omega)$, which is due to the narrow-band noise. For weak narrow-band noise, this peak is described by  Eq.~(4) of the main text if one replaces in this equation $\Gamma$ with $\tilde\Gamma$, $\omega_0$ with $\tilde\omega_0$, and $\Xi(\omega)$ with $\Xi_{\rm nb}(\omega)$. This can be seen from Eq.~(\ref{eq:Phi_F_convenient}). Indeed, in the expression  for $\chi_{\rm sl}(t,t')$ in  Eq.~(\ref{eq:Phi_F_convenient}) one can write $\int_{t'}^t dt'' \xi_{\rm nb}(t'') \approx \xi_{\rm nb}(t)(t-t')$. To find  $\Phi_F^{\rm (nb)}(\omega)$, one should integrate over the range of $t$ given by the reciprocal bandwidth of the narrow-band noise. Since it largely exceeds $1/\Gamma$, the contributions of the broad-band noise to $\chi_{\rm sl}(t,t')$ and $\chi_{\rm sl}(0,t_1')$ are statistically independent. Therefore the averaging over the broad-band noise in these susceptibilities can be done independently. If this averaging is denoted by $\langle\cdot\rangle_{\rm bb}$,
\begin{equation}
\label{eq:auxiliary_for_chi_averaging1}
\langle   \chi_{\rm sl}(t,t')\chi_{\rm sl}^*(0,t_1')\rangle_{\rm bb} \approx \langle   \chi_{\rm sl}(t,t')\rangle_{\rm bb} \langle\chi_{\rm sl}^*(0,t_1') \rangle_{\rm bb}\nonumber
\end{equation}
for $\Gamma t\gg 1$. Here
\begin{equation}
\label{eq:auxiliary_averaging2}
\langle   \chi_{\rm sl}(t,t')\rangle_{\rm bb} \approx e^{-\left[\tilde\Gamma -i\bigl(\omega_F-\tilde\omega_0 -\xi_{\rm nb}(t)\bigl)\right](t-t')}.\nonumber
\end{equation}

Since function $\Xi_{\rm nb}(\omega)$ quickly falls off with increasing $|\omega|$, in the denominator of Eq.~(4) of the main text one can replace $\omega$ with $\omega_F$. One then sees that, to the leading order in the narrow-band noise strength, the ratio of the area $S_{\rm nb}$ of the narrow peak $\Phi_F^{\rm (nb)}(\omega)$  to the area of the $\delta$-peak in the spectrum is
\begin{align}
\label{eq:narrow_band_area}
\frac{S_{\rm nb}}{S_\delta} \approx \frac{1}{2\pi}\,\frac{1}{\tilde\Gamma^2 + (\omega_F- \tilde\omega_0)^2}\int d\omega\, \Xi_{\rm nb}(\omega).
\end{align}

Equations (\ref{eq:area_delta_peak}) and (\ref{eq:narrow_band_area}) can be used to scale the area of the broader peak of $\Phi_F(\omega)$ by $S_\delta$ with account taken of the effect of the narrow-band frequency noise. Along with the onset of a narrow peak in $\Phi_F$, this noise leads to the change of the shape and area of the broad peak. Where the narrow-band noise is weak, the leading-order correction can be found by replacing $\delta\omega_F$ with $\delta\omega_F - \xi_{\rm nb}(t)$ in Eq.~(\ref{eq:Phi_F_convenient}) for $\chi_{\rm sl}(t,t')$ and then expanding to second order in $\xi_{\rm nb}(t)$.  The result is particularly simple in the considered here case where the broad-band frequency noise is $\delta$-correlated in slow time and the broad peak of $\Phi_F(\omega)$ is described by Eq.~(5) of the main text. One just has to replace in this equations $\tilde\omega_0$ with $\tilde\omega_0+\xi_{\rm nb}(t)$, expand in $\xi_{\rm nb}(t)$ to the second order, and average $\xi^2_{\rm nb}(t)\to \langle \xi^2_{\rm nb}(t)\rangle$. The corresponding expression for the area $S_{\rm bb}$ of the broad peak of $\Phi_F(\omega)$ reads
\begin{align}
\label{eq:broad_band_corrected}
S_{\rm bb}&\approx \frac{\pi}{8\omega_0^2}\frac{\Re~\!\mu(1)/\Gamma}{[\tilde\Gamma + (\omega_F - \tilde\omega_0)]^2}\nonumber \\
& \times \left[1-\frac{\tilde\Gamma^2-3(\omega_F-\tilde\omega_0)^2)}{2\pi[\tilde\Gamma^2+(\omega_F-\tilde\omega_0)^2]^2}\int d\omega\, \Xi_{nb}(\omega) \right].
\end{align}
Equations (\ref{eq:narrow_band_area}) and (\ref{eq:broad_band_corrected}) lead to Eq.~(6) of the main text, which shows the contribution of the frequency noise to the width of the broad peak of the spectrum.

\section{Numerical simulations}

The results of the simulations presented in Figs.~2 and 4 of the main text were obtained in a standard way. We integrated the stochastic differential equation (\ref{eq:dot_u}) using the Heun scheme \cite{Mannella2002a}. For a nonlinear oscillator this equation has the extra term $3i(\gamma/2\omega_0) |u|^2u$ in the right-hand side  \cite{DK_review84}. For a nonlinear oscillator, we verified that the values of the modulating field amplitude $F$ were in the range where the driving-induced term in the power spectrum was quadratic in $F$. As seen from the inset in Fig.~4 of the main text, the simulations are in excellent agreement with analytical results  \cite{DK_review84} in the absence of driving. 


\begin{thebibliography}{37}%
\makeatletter
\providecommand \@ifxundefined [1]{%
 \@ifx{#1\undefined}
}%
\providecommand \@ifnum [1]{%
 \ifnum #1\expandafter \@firstoftwo
 \else \expandafter \@secondoftwo
 \fi
}%
\providecommand \@ifx [1]{%
 \ifx #1\expandafter \@firstoftwo
 \else \expandafter \@secondoftwo
 \fi
}%
\providecommand \natexlab [1]{#1}%
\providecommand \enquote  [1]{``#1''}%
\providecommand \bibnamefont  [1]{#1}%
\providecommand \bibfnamefont [1]{#1}%
\providecommand \citenamefont [1]{#1}%
\providecommand \href@noop [0]{\@secondoftwo}%
\providecommand \href [0]{\begingroup \@sanitize@url \@href}%
\providecommand \@href[1]{\@@startlink{#1}\@@href}%
\providecommand \@@href[1]{\endgroup#1\@@endlink}%
\providecommand \@sanitize@url [0]{\catcode `\\12\catcode `\$12\catcode
  `\&12\catcode `\#12\catcode `\^12\catcode `\_12\catcode `\%12\relax}%
\providecommand \@@startlink[1]{}%
\providecommand \@@endlink[0]{}%
\providecommand \url  [0]{\begingroup\@sanitize@url \@url }%
\providecommand \@url [1]{\endgroup\@href {#1}{\urlprefix }}%
\providecommand \urlprefix  [0]{URL }%
\providecommand \Eprint [0]{\href }%
\providecommand \doibase [0]{http://dx.doi.org/}%
\providecommand \selectlanguage [0]{\@gobble}%
\providecommand \bibinfo  [0]{\@secondoftwo}%
\providecommand \bibfield  [0]{\@secondoftwo}%
\providecommand \translation [1]{[#1]}%
\providecommand \BibitemOpen [0]{}%
\providecommand \bibitemStop [0]{}%
\providecommand \bibitemNoStop [0]{.\EOS\space}%
\providecommand \EOS [0]{\spacefactor3000\relax}%
\providecommand \BibitemShut  [1]{\csname bibitem#1\endcsname}%
\let\auto@bib@innerbib\@empty
\bibitem [{\citenamefont {Lorentz}(1916)}]{Lorentz1916}%
  \BibitemOpen
  \bibfield  {author} {\bibinfo {author} {\bibfnamefont {H.~A.}\ \bibnamefont
  {Lorentz}},\ }\href@noop {} {\emph {\bibinfo {title} {The theory of electrons
  and its applications to the phenomena of light and radiant heat}}}\ (\bibinfo
   {publisher} {Teubner, B. G.},\ \bibinfo {year} {Leipzig, 1916})\BibitemShut
  {NoStop}%
\bibitem [{\citenamefont {Einstein}\ and\ \citenamefont
  {Hopf}(1910)}]{Einstein1910b}%
  \BibitemOpen
  \bibfield  {author} {\bibinfo {author} {\bibfnamefont {A.}~\bibnamefont
  {Einstein}}\ and\ \bibinfo {author} {\bibfnamefont {L.}~\bibnamefont
  {Hopf}},\ }\href@noop {} {\bibfield  {journal} {\bibinfo  {journal} {Ann.d.
  Phys.}\ }\textbf {\bibinfo {volume} {33}},\ \bibinfo {pages} {1105} (\bibinfo
  {year} {1910})}\BibitemShut {NoStop}%
\bibitem [{\citenamefont {Heitler}(2010)}]{Heitler2010}%
  \BibitemOpen
  \bibfield  {author} {\bibinfo {author} {\bibfnamefont {W.}~\bibnamefont
  {Heitler}},\ }\href@noop {} {\emph {\bibinfo {title} {The Quantum Theory of
  Radiation, 3rd ed.}}}\ (\bibinfo  {publisher} {Dover Publications, Inc., New
  York},\ \bibinfo {year} {2010})\BibitemShut {NoStop}%
\bibitem [{\citenamefont {Cleland}\ and\ \citenamefont
  {Roukes}(2002)}]{Cleland2002}%
  \BibitemOpen
  \bibfield  {author} {\bibinfo {author} {\bibfnamefont {A.~N.}\ \bibnamefont
  {Cleland}}\ and\ \bibinfo {author} {\bibfnamefont {M.~L.}\ \bibnamefont
  {Roukes}},\ }\href@noop {} {\bibfield  {journal} {\bibinfo  {journal} {J.
  Appl. Phys.}\ }\textbf {\bibinfo {volume} {92}},\ \bibinfo {pages} {2758}
  (\bibinfo {year} {2002})}\BibitemShut {NoStop}%
\bibitem [{\citenamefont {Sazonova}\ \emph {et~al.}(2004)\citenamefont
  {Sazonova}, \citenamefont {Yaish}, \citenamefont {Ustunel}, \citenamefont
  {Roundy}, \citenamefont {Arias},\ and\ \citenamefont
  {McEuen}}]{Sazonova2004}%
  \BibitemOpen
  \bibfield  {author} {\bibinfo {author} {\bibfnamefont {V.}~\bibnamefont
  {Sazonova}}, \bibinfo {author} {\bibfnamefont {Y.}~\bibnamefont {Yaish}},
  \bibinfo {author} {\bibfnamefont {H.}~\bibnamefont {Ustunel}}, \bibinfo
  {author} {\bibfnamefont {D.}~\bibnamefont {Roundy}}, \bibinfo {author}
  {\bibfnamefont {T.~A.}\ \bibnamefont {Arias}}, \ and\ \bibinfo {author}
  {\bibfnamefont {P.~L.}\ \bibnamefont {McEuen}},\ }\href@noop {} {\bibfield
  {journal} {\bibinfo  {journal} {Nature}\ }\textbf {\bibinfo {volume} {431}},\
  \bibinfo {pages} {284} (\bibinfo {year} {2004})}\BibitemShut {NoStop}%
\bibitem [{\citenamefont {Lifshitz}\ and\ \citenamefont
  {Cross}(2008)}]{Lifshitz2008}%
  \BibitemOpen
  \bibfield  {author} {\bibinfo {author} {\bibfnamefont {R.}~\bibnamefont
  {Lifshitz}}\ and\ \bibinfo {author} {\bibfnamefont {M.~C.}\ \bibnamefont
  {Cross}},\ }in\ \href@noop {} {\emph {\bibinfo {booktitle} {Review of
  Nonlinear Dynamics and Complexity}}},\ \bibinfo {editor} {edited by\ \bibinfo
  {editor} {\bibfnamefont {H.~G.}\ \bibnamefont {Schuster}}}\ (\bibinfo
  {publisher} {Wiley, Weinheim},\ \bibinfo {year} {2008})\ pp.\ \bibinfo
  {pages} {1--52}\BibitemShut {NoStop}%
\bibitem [{\citenamefont {Steele}\ \emph {et~al.}(2009)\citenamefont {Steele},
  \citenamefont {Huttel}, \citenamefont {Witkamp}, \citenamefont {Poot},
  \citenamefont {Meerwaldt}, \citenamefont {Kouwenhoven},\ and\ \citenamefont
  {van~der Zant}}]{Steele2009}%
  \BibitemOpen
  \bibfield  {author} {\bibinfo {author} {\bibfnamefont {G.~A.}\ \bibnamefont
  {Steele}}, \bibinfo {author} {\bibfnamefont {A.~K.}\ \bibnamefont {Huttel}},
  \bibinfo {author} {\bibfnamefont {B.}~\bibnamefont {Witkamp}}, \bibinfo
  {author} {\bibfnamefont {M.}~\bibnamefont {Poot}}, \bibinfo {author}
  {\bibfnamefont {H.~B.}\ \bibnamefont {Meerwaldt}}, \bibinfo {author}
  {\bibfnamefont {L.~P.}\ \bibnamefont {Kouwenhoven}}, \ and\ \bibinfo {author}
  {\bibfnamefont {H.~S.~J.}\ \bibnamefont {van~der Zant}},\ }\href@noop {}
  {\bibfield  {journal} {\bibinfo  {journal} {Science}\ }\textbf {\bibinfo
  {volume} {325}},\ \bibinfo {pages} {1103} (\bibinfo {year}
  {2009})}\BibitemShut {NoStop}%
\bibitem [{\citenamefont {Lassagne}\ \emph {et~al.}(2009)\citenamefont
  {Lassagne}, \citenamefont {Tarakanov}, \citenamefont {Kinaret}, \citenamefont
  {Garcia-Sanchez},\ and\ \citenamefont {Bachtold}}]{Lassagne2009}%
  \BibitemOpen
  \bibfield  {author} {\bibinfo {author} {\bibfnamefont {B.}~\bibnamefont
  {Lassagne}}, \bibinfo {author} {\bibfnamefont {Y.}~\bibnamefont {Tarakanov}},
  \bibinfo {author} {\bibfnamefont {J.}~\bibnamefont {Kinaret}}, \bibinfo
  {author} {\bibfnamefont {D.}~\bibnamefont {Garcia-Sanchez}}, \ and\ \bibinfo
  {author} {\bibfnamefont {A.}~\bibnamefont {Bachtold}},\ }\href@noop {}
  {\bibfield  {journal} {\bibinfo  {journal} {Science}\ }\textbf {\bibinfo
  {volume} {325}},\ \bibinfo {pages} {1107} (\bibinfo {year}
  {2009})}\BibitemShut {NoStop}%
\bibitem [{\citenamefont {Fong}\ \emph {et~al.}(2012)\citenamefont {Fong},
  \citenamefont {Pernice},\ and\ \citenamefont {Tang}}]{Fong2012}%
  \BibitemOpen
  \bibfield  {author} {\bibinfo {author} {\bibfnamefont {K.~Y.}\ \bibnamefont
  {Fong}}, \bibinfo {author} {\bibfnamefont {W.~H.~P.}\ \bibnamefont
  {Pernice}}, \ and\ \bibinfo {author} {\bibfnamefont {H.~X.}\ \bibnamefont
  {Tang}},\ }\href@noop {} {\bibfield  {journal} {\bibinfo  {journal} {Phys.
  Rev. B}\ }\textbf {\bibinfo {volume} {85}},\ \bibinfo {pages} {161410 (R)}
  (\bibinfo {year} {2012})}\BibitemShut {NoStop}%
\bibitem [{\citenamefont {Siria}\ \emph {et~al.}(2012)\citenamefont {Siria},
  \citenamefont {Barois}, \citenamefont {Vilella}, \citenamefont {Perisanu},
  \citenamefont {Ayari}, \citenamefont {Guillot}, \citenamefont {Purcell},\
  and\ \citenamefont {Poncharal}}]{Siria2012}%
  \BibitemOpen
  \bibfield  {author} {\bibinfo {author} {\bibfnamefont {A.}~\bibnamefont
  {Siria}}, \bibinfo {author} {\bibfnamefont {T.}~\bibnamefont {Barois}},
  \bibinfo {author} {\bibfnamefont {K.}~\bibnamefont {Vilella}}, \bibinfo
  {author} {\bibfnamefont {S.}~\bibnamefont {Perisanu}}, \bibinfo {author}
  {\bibfnamefont {A.}~\bibnamefont {Ayari}}, \bibinfo {author} {\bibfnamefont
  {D.}~\bibnamefont {Guillot}}, \bibinfo {author} {\bibfnamefont
  {S.}~\bibnamefont {Purcell}}, \ and\ \bibinfo {author} {\bibfnamefont
  {P.}~\bibnamefont {Poncharal}},\ }\href@noop {} {\bibfield  {journal}
  {\bibinfo  {journal} {Nano Lett.}\ }\textbf {\bibinfo {volume} {12}},\
  \bibinfo {pages} {3551} (\bibinfo {year} {2012})}\BibitemShut {NoStop}%
\bibitem [{\citenamefont {Gavartin}\ \emph {et~al.}(2013)\citenamefont
  {Gavartin}, \citenamefont {Verlot},\ and\ \citenamefont
  {Kippenberg}}]{Gavartin2013}%
  \BibitemOpen
  \bibfield  {author} {\bibinfo {author} {\bibfnamefont {E.}~\bibnamefont
  {Gavartin}}, \bibinfo {author} {\bibfnamefont {P.}~\bibnamefont {Verlot}}, \
  and\ \bibinfo {author} {\bibfnamefont {T.~J.}\ \bibnamefont {Kippenberg}},\
  }\href@noop {} {\bibfield  {journal} {\bibinfo  {journal} {Nat. Commun.}\
  }\textbf {\bibinfo {volume} {4}},\ \bibinfo {pages} {2860} (\bibinfo {year}
  {2013})}\BibitemShut {NoStop}%
\bibitem [{\citenamefont {Gao}\ \emph {et~al.}(2007)\citenamefont {Gao},
  \citenamefont {Zmuidzinas}, \citenamefont {Mazin}, \citenamefont {LeDuc},\
  and\ \citenamefont {Day}}]{Gao2007}%
  \BibitemOpen
  \bibfield  {author} {\bibinfo {author} {\bibfnamefont {J.}~\bibnamefont
  {Gao}}, \bibinfo {author} {\bibfnamefont {J.}~\bibnamefont {Zmuidzinas}},
  \bibinfo {author} {\bibfnamefont {B.~A.}\ \bibnamefont {Mazin}}, \bibinfo
  {author} {\bibfnamefont {H.~G.}\ \bibnamefont {LeDuc}}, \ and\ \bibinfo
  {author} {\bibfnamefont {P.~K.}\ \bibnamefont {Day}},\ }\href@noop {}
  {\bibfield  {journal} {\bibinfo  {journal} {Appl. Phys. Lett.}\ }\textbf
  {\bibinfo {volume} {90}},\ \bibinfo {eid} {102507} (\bibinfo {year}
  {2007})}\BibitemShut {NoStop}%
\bibitem [{\citenamefont {Neill}\ \emph {et~al.}(2013)\citenamefont {Neill},
  \citenamefont {Megrant}, \citenamefont {Barends}, \citenamefont {Chen},
  \citenamefont {Chiaro}, \citenamefont {Kelly}, \citenamefont {Mutus},
  \citenamefont {O'Malley}, \citenamefont {Sank}, \citenamefont {Wenner},
  \citenamefont {White}, \citenamefont {Yin}, \citenamefont {Cleland},\ and\
  \citenamefont {Martinis}}]{Neill2013}%
  \BibitemOpen
  \bibfield  {author} {\bibinfo {author} {\bibfnamefont {C.}~\bibnamefont
  {Neill}}, \bibinfo {author} {\bibfnamefont {A.}~\bibnamefont {Megrant}},
  \bibinfo {author} {\bibfnamefont {R.}~\bibnamefont {Barends}}, \bibinfo
  {author} {\bibfnamefont {Y.}~\bibnamefont {Chen}}, \bibinfo {author}
  {\bibfnamefont {B.}~\bibnamefont {Chiaro}}, \bibinfo {author} {\bibfnamefont
  {J.}~\bibnamefont {Kelly}}, \bibinfo {author} {\bibfnamefont {J.~Y.}\
  \bibnamefont {Mutus}}, \bibinfo {author} {\bibfnamefont {P.~J.~J.}\
  \bibnamefont {O'Malley}}, \bibinfo {author} {\bibfnamefont {D.}~\bibnamefont
  {Sank}}, \bibinfo {author} {\bibfnamefont {J.}~\bibnamefont {Wenner}},
  \bibinfo {author} {\bibfnamefont {T.~C.}\ \bibnamefont {White}}, \bibinfo
  {author} {\bibfnamefont {Y.}~\bibnamefont {Yin}}, \bibinfo {author}
  {\bibfnamefont {A.~N.}\ \bibnamefont {Cleland}}, \ and\ \bibinfo {author}
  {\bibfnamefont {J.~M.}\ \bibnamefont {Martinis}},\ }\href@noop {} {\bibfield
  {journal} {\bibinfo  {journal} {Appl. Phys. Lett.}\ }\textbf {\bibinfo
  {volume} {103}},\ \bibinfo {pages} {072601} (\bibinfo {year}
  {2013})}\BibitemShut {NoStop}%
\bibitem [{\citenamefont {Burnett}\ \emph {et~al.}(2014)\citenamefont
  {Burnett}, \citenamefont {Faoro}, \citenamefont {Wisby}, \citenamefont
  {Gurtovoi}, \citenamefont {Chernykh}, \citenamefont {Mikhailov},
  \citenamefont {Tulin}, \citenamefont {Shaikhaidarov}, \citenamefont
  {Antonov}, \citenamefont {Meeson}, \citenamefont {Tzalenchuk},\ and\
  \citenamefont {Lindstrom}}]{Burnett2013a}%
  \BibitemOpen
  \bibfield  {author} {\bibinfo {author} {\bibfnamefont {J.}~\bibnamefont
  {Burnett}}, \bibinfo {author} {\bibfnamefont {L.}~\bibnamefont {Faoro}},
  \bibinfo {author} {\bibfnamefont {I.}~\bibnamefont {Wisby}}, \bibinfo
  {author} {\bibfnamefont {V.~L.}\ \bibnamefont {Gurtovoi}}, \bibinfo {author}
  {\bibfnamefont {A.~V.}\ \bibnamefont {Chernykh}}, \bibinfo {author}
  {\bibfnamefont {G.~M.}\ \bibnamefont {Mikhailov}}, \bibinfo {author}
  {\bibfnamefont {V.~A.}\ \bibnamefont {Tulin}}, \bibinfo {author}
  {\bibfnamefont {R.}~\bibnamefont {Shaikhaidarov}}, \bibinfo {author}
  {\bibfnamefont {V.}~\bibnamefont {Antonov}}, \bibinfo {author} {\bibfnamefont
  {P.~J.}\ \bibnamefont {Meeson}}, \bibinfo {author} {\bibfnamefont {A.~Y.}\
  \bibnamefont {Tzalenchuk}}, \ and\ \bibinfo {author} {\bibfnamefont
  {T.}~\bibnamefont {Lindstrom}},\ }\href@noop {} {\bibfield  {journal}
  {\bibinfo  {journal} {Nat. Commun.}\ }\textbf {\bibinfo {volume} {5}},\
  \bibinfo {pages} {4119} (\bibinfo {year} {2014})}\BibitemShut {NoStop}%
\bibitem [{\citenamefont {{Faoro}}\ and\ \citenamefont
  {{Ioffe}}()}]{Faoro2014}%
  \BibitemOpen
  \bibfield  {author} {\bibinfo {author} {\bibfnamefont {L.}~\bibnamefont
  {{Faoro}}}\ and\ \bibinfo {author} {\bibfnamefont {L.~B.}\ \bibnamefont
  {{Ioffe}}},\ }\href@noop {} {\ }\Eprint {http://arxiv.org/abs/1404.2410}
  {arXiv:1404.2410} \BibitemShut {NoStop}%
\bibitem [{\citenamefont {Aspelmeyer}\ \emph {et~al.}(2014)\citenamefont
  {Aspelmeyer}, \citenamefont {Kippenberg},\ and\ \citenamefont
  {Marquardt}}]{Aspelmeyer2013}%
  \BibitemOpen
  \bibfield  {author} {\bibinfo {author} {\bibfnamefont {M.}~\bibnamefont
  {Aspelmeyer}}, \bibinfo {author} {\bibfnamefont {T.~J.}\ \bibnamefont
  {Kippenberg}}, \ and\ \bibinfo {author} {\bibfnamefont {F.}~\bibnamefont
  {Marquardt}},\ }\href@noop {} {\bibfield  {journal} {\bibinfo  {journal}
  {Rev. Mod. Phys.}\ ,\ \bibinfo {pages} {to be published}}  (\bibinfo {year}
  {2014})}\BibitemShut {NoStop}%
\bibitem [{\citenamefont {Dykman}\ \emph {et~al.}(2010)\citenamefont {Dykman},
  \citenamefont {Khasin}, \citenamefont {Portman},\ and\ \citenamefont
  {Shaw}}]{Dykman2010}%
  \BibitemOpen
  \bibfield  {author} {\bibinfo {author} {\bibfnamefont {M.~I.}\ \bibnamefont
  {Dykman}}, \bibinfo {author} {\bibfnamefont {M.}~\bibnamefont {Khasin}},
  \bibinfo {author} {\bibfnamefont {J.}~\bibnamefont {Portman}}, \ and\
  \bibinfo {author} {\bibfnamefont {S.~W.}\ \bibnamefont {Shaw}},\ }\href@noop
  {} {\bibfield  {journal} {\bibinfo  {journal} {Phys. Rev. Lett.}\ }\textbf
  {\bibinfo {volume} {105}},\ \bibinfo {pages} {230601} (\bibinfo {year}
  {2010})}\BibitemShut {NoStop}%
\bibitem [{\citenamefont {Yang}\ \emph {et~al.}(2011)\citenamefont {Yang},
  \citenamefont {Callegari}, \citenamefont {Feng},\ and\ \citenamefont
  {Roukes}}]{Yang2011}%
  \BibitemOpen
  \bibfield  {author} {\bibinfo {author} {\bibfnamefont {Y.~T.}\ \bibnamefont
  {Yang}}, \bibinfo {author} {\bibfnamefont {C.}~\bibnamefont {Callegari}},
  \bibinfo {author} {\bibfnamefont {X.~L.}\ \bibnamefont {Feng}}, \ and\
  \bibinfo {author} {\bibfnamefont {M.~L.}\ \bibnamefont {Roukes}},\
  }\href@noop {} {\bibfield  {journal} {\bibinfo  {journal} {Nano Lett.}\
  }\textbf {\bibinfo {volume} {11}},\ \bibinfo {pages} {1753} (\bibinfo {year}
  {2011})}\BibitemShut {NoStop}%
\bibitem [{\citenamefont {Barnard}\ \emph {et~al.}(2012)\citenamefont
  {Barnard}, \citenamefont {Sazonova}, \citenamefont {van~der Zande},\ and\
  \citenamefont {McEuen}}]{Barnard2012}%
  \BibitemOpen
  \bibfield  {author} {\bibinfo {author} {\bibfnamefont {A.~W.}\ \bibnamefont
  {Barnard}}, \bibinfo {author} {\bibfnamefont {V.}~\bibnamefont {Sazonova}},
  \bibinfo {author} {\bibfnamefont {A.~M.}\ \bibnamefont {van~der Zande}}, \
  and\ \bibinfo {author} {\bibfnamefont {P.~L.}\ \bibnamefont {McEuen}},\
  }\href@noop {} {\bibfield  {journal} {\bibinfo  {journal} {PNAS}\ }\textbf
  {\bibinfo {volume} {109}},\ \bibinfo {pages} {19093} (\bibinfo {year}
  {2012})}\BibitemShut {NoStop}%
\bibitem [{\citenamefont {Miao}\ \emph {et~al.}(2014)\citenamefont {Miao},
  \citenamefont {Yeom}, \citenamefont {Wang}, \citenamefont {Standley},\ and\
  \citenamefont {Bockrath}}]{Miao2014}%
  \BibitemOpen
  \bibfield  {author} {\bibinfo {author} {\bibfnamefont {T.~F.}\ \bibnamefont
  {Miao}}, \bibinfo {author} {\bibfnamefont {S.}~\bibnamefont {Yeom}}, \bibinfo
  {author} {\bibfnamefont {P.}~\bibnamefont {Wang}}, \bibinfo {author}
  {\bibfnamefont {B.}~\bibnamefont {Standley}}, \ and\ \bibinfo {author}
  {\bibfnamefont {M.}~\bibnamefont {Bockrath}},\ }\href {\doibase
  10.1021/nl403936a} {\bibfield  {journal} {\bibinfo  {journal} {Nano Lett.}\
  }\textbf {\bibinfo {volume} {14}},\ \bibinfo {pages} {2982} (\bibinfo {year}
  {2014})}\BibitemShut {NoStop}%
\bibitem [{\citenamefont {Eichler}\ \emph {et~al.}(2013)\citenamefont
  {Eichler}, \citenamefont {Moser}, \citenamefont {Dykman},\ and\ \citenamefont
  {Bachtold}}]{Eichler2013}%
  \BibitemOpen
  \bibfield  {author} {\bibinfo {author} {\bibfnamefont {A.}~\bibnamefont
  {Eichler}}, \bibinfo {author} {\bibfnamefont {J.}~\bibnamefont {Moser}},
  \bibinfo {author} {\bibfnamefont {M.~I.}\ \bibnamefont {Dykman}}, \ and\
  \bibinfo {author} {\bibfnamefont {A.}~\bibnamefont {Bachtold}},\ }\href@noop
  {} {\bibfield  {journal} {\bibinfo  {journal} {Nat. Commun.}\ }\textbf
  {\bibinfo {volume} {4}},\ \bibinfo {pages} {2843} (\bibinfo {year}
  {2013})}\BibitemShut {NoStop}%
\bibitem [{\citenamefont {{Meenehan}}\ \emph {et~al.}(2014)\citenamefont
  {{Meenehan}}, \citenamefont {{Cohen}}, \citenamefont {{Groeblacher}},
  \citenamefont {{Hill}}, \citenamefont {{Safavi-Naeini}}, \citenamefont
  {{Aspelmeyer}},\ and\ \citenamefont {{Painter}}}]{Meenehan2014}%
  \BibitemOpen
  \bibfield  {author} {\bibinfo {author} {\bibfnamefont {S.~M.}\ \bibnamefont
  {{Meenehan}}}, \bibinfo {author} {\bibfnamefont {J.~D.}\ \bibnamefont
  {{Cohen}}}, \bibinfo {author} {\bibfnamefont {S.}~\bibnamefont
  {{Groeblacher}}}, \bibinfo {author} {\bibfnamefont {J.~T.}\ \bibnamefont
  {{Hill}}}, \bibinfo {author} {\bibfnamefont {A.~H.}\ \bibnamefont
  {{Safavi-Naeini}}}, \bibinfo {author} {\bibfnamefont {M.}~\bibnamefont
  {{Aspelmeyer}}}, \ and\ \bibinfo {author} {\bibfnamefont {O.}~\bibnamefont
  {{Painter}}},\ }\href@noop {} {\bibfield  {journal} {\bibinfo  {journal}
  {Phys. Rev. A}\ }\textbf {\bibinfo {volume} {90}},\ \bibinfo {pages} {011803}
  (\bibinfo {year} {2014})}\BibitemShut {NoStop}%
\bibitem [{\citenamefont {Einstein}(1910)}]{Einstein1910}%
  \BibitemOpen
  \bibfield  {author} {\bibinfo {author} {\bibfnamefont {A.}~\bibnamefont
  {Einstein}},\ }\href@noop {} {\bibfield  {journal} {\bibinfo  {journal} {Ann.
  d. Phys.}\ }\textbf {\bibinfo {volume} {33}},\ \bibinfo {pages} {1275}
  (\bibinfo {year} {1910})}\BibitemShut {NoStop}%
\bibitem [{\citenamefont {Lindenberg}\ \emph {et~al.}(1981)\citenamefont
  {Lindenberg}, \citenamefont {Seshadri},\ and\ \citenamefont
  {West}}]{Lindenberg1981}%
  \BibitemOpen
  \bibfield  {author} {\bibinfo {author} {\bibfnamefont {K.}~\bibnamefont
  {Lindenberg}}, \bibinfo {author} {\bibfnamefont {V.}~\bibnamefont
  {Seshadri}}, \ and\ \bibinfo {author} {\bibfnamefont {B.~J.}\ \bibnamefont
  {West}},\ }\href@noop {} {\bibfield  {journal} {\bibinfo  {journal} {Physica
  A}\ }\textbf {\bibinfo {volume} {105}},\ \bibinfo {pages} {445} (\bibinfo
  {year} {1981})}\BibitemShut {NoStop}%
\bibitem [{\citenamefont {Gitterman}(2005)}]{Gitterman_book2005}%
  \BibitemOpen
  \bibfield  {author} {\bibinfo {author} {\bibfnamefont {M.}~\bibnamefont
  {Gitterman}},\ }\href@noop {} {\emph {\bibinfo {title} {The Noisy
  Oscillator}}}\ (\bibinfo  {publisher} {World Scientific},\ \bibinfo {year}
  {New Jersey, 2005})\BibitemShut {NoStop}%
\bibitem [{\citenamefont {Maizelis}\ \emph {et~al.}(2011)\citenamefont
  {Maizelis}, \citenamefont {Roukes},\ and\ \citenamefont
  {Dykman}}]{Maizelis2011}%
  \BibitemOpen
  \bibfield  {author} {\bibinfo {author} {\bibfnamefont {Z.~A.}\ \bibnamefont
  {Maizelis}}, \bibinfo {author} {\bibfnamefont {M.~L.}\ \bibnamefont
  {Roukes}}, \ and\ \bibinfo {author} {\bibfnamefont {M.~I.}\ \bibnamefont
  {Dykman}},\ }\href@noop {} {\bibfield  {journal} {\bibinfo  {journal} {Phys.
  Rev. B}\ }\textbf {\bibinfo {volume} {84}},\ \bibinfo {pages} {144301}
  (\bibinfo {year} {2011})}\BibitemShut {NoStop}%
\bibitem [{\citenamefont {Moser}\ \emph {et~al.}(2013)\citenamefont {Moser},
  \citenamefont {G\"uttinger}, \citenamefont {Eichler}, \citenamefont
  {Esplandiu}, \citenamefont {Liu}, \citenamefont {Dykman},\ and\ \citenamefont
  {Bachtold}}]{Moser2013}%
  \BibitemOpen
  \bibfield  {author} {\bibinfo {author} {\bibfnamefont {J.}~\bibnamefont
  {Moser}}, \bibinfo {author} {\bibfnamefont {J.}~\bibnamefont {G\"uttinger}},
  \bibinfo {author} {\bibfnamefont {A.}~\bibnamefont {Eichler}}, \bibinfo
  {author} {\bibfnamefont {M.~J.}\ \bibnamefont {Esplandiu}}, \bibinfo {author}
  {\bibfnamefont {D.~E.}\ \bibnamefont {Liu}}, \bibinfo {author} {\bibfnamefont
  {M.~I.}\ \bibnamefont {Dykman}}, \ and\ \bibinfo {author} {\bibfnamefont
  {A.}~\bibnamefont {Bachtold}},\ }\href@noop {} {\bibfield  {journal}
  {\bibinfo  {journal} {Nat. Nanotech.}\ }\textbf {\bibinfo {volume} {8}},\
  \bibinfo {pages} {493} (\bibinfo {year} {2013})}\BibitemShut {NoStop}%
\bibitem [{\citenamefont {Dykman}(2012)}]{Dykman2012b}%
  \BibitemOpen
  \bibinfo {editor} {\bibfnamefont {M.~I.}\ \bibnamefont {Dykman}},\ ed.,\
  \href@noop {} {\emph {\bibinfo {title} {Fluctuating Nonlinear Oscillators:
  from Nanomechanics to Quantum Superconducting Circuits}}}\ (\bibinfo
  {publisher} {OUP, Oxford},\ \bibinfo {year} {2012})\BibitemShut {NoStop}%
\bibitem [{\citenamefont {Dykman}\ and\ \citenamefont
  {Krivoglaz}(1984)}]{DK_review84}%
  \BibitemOpen
  \bibfield  {author} {\bibinfo {author} {\bibfnamefont {M.~I.}\ \bibnamefont
  {Dykman}}\ and\ \bibinfo {author} {\bibfnamefont {M.~A.}\ \bibnamefont
  {Krivoglaz}},\ }in\ \href@noop {} {\emph {\bibinfo {booktitle} {Sov. Phys.
  Reviews}}},\ Vol.~\bibinfo {volume} {5},\ \bibinfo {editor} {edited by\
  \bibinfo {editor} {\bibfnamefont {I.~M.}\ \bibnamefont {Khalatnikov}}}\
  (\bibinfo  {publisher} {Harwood Academic, New York},\ \bibinfo {year}
  {1984})\ pp.\ \bibinfo {pages} {265--441}\BibitemShut {NoStop}%
\bibitem [{\citenamefont {Blencowe}(2004)}]{Blencowe2004a}%
  \BibitemOpen
  \bibfield  {author} {\bibinfo {author} {\bibfnamefont {M.}~\bibnamefont
  {Blencowe}},\ }\href@noop {} {\bibfield  {journal} {\bibinfo  {journal}
  {Phys. Rep.}\ }\textbf {\bibinfo {volume} {395}},\ \bibinfo {pages} {159}
  (\bibinfo {year} {2004})}\BibitemShut {NoStop}%
\bibitem [{\citenamefont {Schwab}\ and\ \citenamefont
  {Roukes}(2005)}]{Schwab2005a}%
  \BibitemOpen
  \bibfield  {author} {\bibinfo {author} {\bibfnamefont {K.~C.}\ \bibnamefont
  {Schwab}}\ and\ \bibinfo {author} {\bibfnamefont {M.~L.}\ \bibnamefont
  {Roukes}},\ }\href@noop {} {\bibfield  {journal} {\bibinfo  {journal} {Phys.
  Today}\ }\textbf {\bibinfo {volume} {58}},\ \bibinfo {pages} {36} (\bibinfo
  {year} {2005})}\BibitemShut {NoStop}%
\bibitem [{\citenamefont {Clerk}\ \emph {et~al.}(2010)\citenamefont {Clerk},
  \citenamefont {Marquardt},\ and\ \citenamefont {Harris}}]{Clerk2010a}%
  \BibitemOpen
  \bibfield  {author} {\bibinfo {author} {\bibfnamefont {A.~A.}\ \bibnamefont
  {Clerk}}, \bibinfo {author} {\bibfnamefont {F.}~\bibnamefont {Marquardt}}, \
  and\ \bibinfo {author} {\bibfnamefont {J.~G.~E.}\ \bibnamefont {Harris}},\
  }\href@noop {} {\bibfield  {journal} {\bibinfo  {journal} {Phys. Rev. Lett.}\
  }\textbf {\bibinfo {volume} {104}},\ \bibinfo {pages} {213603} (\bibinfo
  {year} {2010})}\BibitemShut {NoStop}%
\bibitem [{\citenamefont {O'Connell}\ \emph {et~al.}(2010)\citenamefont
  {O'Connell}, \citenamefont {Hofheinz}, \citenamefont {Ansmann}, \citenamefont
  {Bialczak}, \citenamefont {Lenander}, \citenamefont {Lucero}, \citenamefont
  {Neeley}, \citenamefont {Sank}, \citenamefont {Wang}, \citenamefont {Weides},
  \citenamefont {Wenner}, \citenamefont {Martinis},\ and\ \citenamefont
  {Cleland}}]{O'Connell2010}%
  \BibitemOpen
  \bibfield  {author} {\bibinfo {author} {\bibfnamefont {A.~D.}\ \bibnamefont
  {O'Connell}}, \bibinfo {author} {\bibfnamefont {M.}~\bibnamefont {Hofheinz}},
  \bibinfo {author} {\bibfnamefont {M.}~\bibnamefont {Ansmann}}, \bibinfo
  {author} {\bibfnamefont {R.~C.}\ \bibnamefont {Bialczak}}, \bibinfo {author}
  {\bibfnamefont {M.}~\bibnamefont {Lenander}}, \bibinfo {author}
  {\bibfnamefont {E.}~\bibnamefont {Lucero}}, \bibinfo {author} {\bibfnamefont
  {M.}~\bibnamefont {Neeley}}, \bibinfo {author} {\bibfnamefont
  {D.}~\bibnamefont {Sank}}, \bibinfo {author} {\bibfnamefont {H.}~\bibnamefont
  {Wang}}, \bibinfo {author} {\bibfnamefont {M.}~\bibnamefont {Weides}},
  \bibinfo {author} {\bibfnamefont {J.}~\bibnamefont {Wenner}}, \bibinfo
  {author} {\bibfnamefont {J.~M.}\ \bibnamefont {Martinis}}, \ and\ \bibinfo
  {author} {\bibfnamefont {A.~N.}\ \bibnamefont {Cleland}},\ }\href {\doibase
  10.1038/nature08967} {\bibfield  {journal} {\bibinfo  {journal} {Nature}\
  }\textbf {\bibinfo {volume} {464}},\ \bibinfo {pages} {697} (\bibinfo {year}
  {2010})}\BibitemShut {NoStop}%
\bibitem [{\citenamefont {Feynman}\ and\ \citenamefont
  {Hibbs}(1965)}]{FeynmanQM}%
  \BibitemOpen
  \bibfield  {author} {\bibinfo {author} {\bibfnamefont {R.~P.}\ \bibnamefont
  {Feynman}}\ and\ \bibinfo {author} {\bibfnamefont {A.~R.}\ \bibnamefont
  {Hibbs}},\ }\href@noop {} {\emph {\bibinfo {title} {Quantum Mechanics and
  Path Integrals}}}\ (\bibinfo  {publisher} {McGraw-Hill},\ \bibinfo {address}
  {New-York},\ \bibinfo {year} {1965})\BibitemShut {NoStop}%
\bibitem [{\citenamefont {Kim}\ \emph {et~al.}(2001)\citenamefont {Kim},
  \citenamefont {Shi}, \citenamefont {Majumdar},\ and\ \citenamefont
  {McEuen}}]{Kim2001}%
  \BibitemOpen
  \bibfield  {author} {\bibinfo {author} {\bibfnamefont {P.}~\bibnamefont
  {Kim}}, \bibinfo {author} {\bibfnamefont {L.}~\bibnamefont {Shi}}, \bibinfo
  {author} {\bibfnamefont {A.}~\bibnamefont {Majumdar}}, \ and\ \bibinfo
  {author} {\bibfnamefont {P.}~\bibnamefont {McEuen}},\ }\href {\doibase
  10.1103/PhysRevLett.87.215502} {\bibfield  {journal} {\bibinfo  {journal}
  {Phys. Rev. Lett.}\ }\textbf {\bibinfo {volume} {87}},\ \bibinfo {pages}
  {215502} (\bibinfo {year} {2001})}\BibitemShut {NoStop}%
\bibitem [{\citenamefont {Hone}\ \emph {et~al.}(2000)\citenamefont {Hone},
  \citenamefont {Llaguno}, \citenamefont {Nemes}, \citenamefont {Johnson},
  \citenamefont {Fischer}, \citenamefont {Walters}, \citenamefont {Casavant},
  \citenamefont {Schmidt},\ and\ \citenamefont {Smalley}}]{Hone2000}%
  \BibitemOpen
  \bibfield  {author} {\bibinfo {author} {\bibfnamefont {J.}~\bibnamefont
  {Hone}}, \bibinfo {author} {\bibfnamefont {M.}~\bibnamefont {Llaguno}},
  \bibinfo {author} {\bibfnamefont {N.}~\bibnamefont {Nemes}}, \bibinfo
  {author} {\bibfnamefont {A.}~\bibnamefont {Johnson}}, \bibinfo {author}
  {\bibfnamefont {J.}~\bibnamefont {Fischer}}, \bibinfo {author} {\bibfnamefont
  {D.}~\bibnamefont {Walters}}, \bibinfo {author} {\bibfnamefont
  {M.}~\bibnamefont {Casavant}}, \bibinfo {author} {\bibfnamefont
  {J.}~\bibnamefont {Schmidt}}, \ and\ \bibinfo {author} {\bibfnamefont
  {R.}~\bibnamefont {Smalley}},\ }\href {\doibase 10.1063/1.127079} {\bibfield
  {journal} {\bibinfo  {journal} {Appl. Phys. Lett.}\ }\textbf {\bibinfo
  {volume} {77}},\ \bibinfo {pages} {666} (\bibinfo {year} {2000})}\BibitemShut
  {NoStop}%
\bibitem [{\citenamefont {Mannella}(2002)}]{Mannella2002a}%
  \BibitemOpen
  \bibfield  {author} {\bibinfo {author} {\bibfnamefont {R.}~\bibnamefont
  {Mannella}},\ }\href@noop {} {\bibfield  {journal} {\bibinfo  {journal} {Int.
  J. Mod. Phys. C}\ }\textbf {\bibinfo {volume} {13}},\ \bibinfo {pages} {1177}
  (\bibinfo {year} {2002})}\BibitemShut {NoStop}%
\end{thebibliography}


%
\end{document}